\documentclass{SCIS2024}
\pdfoutput=1 

\newcommand{\ba}{\begin{array}}
\newcommand{\ea}{\end{array}}
\newcommand{\be}{\begin{displaymath}}
\newcommand{\ee}{\end{displaymath}}
\newcommand{\ben}{\begin{equation}}
\newcommand{\een}{\end{equation}}
\newcommand{\bena}{\begin{eqnarray}}
\newcommand{\eena}{\end{eqnarray}}
\newcommand{\beqa}{\begin{eqnarray*}}
\newcommand{\enqa}{\end{eqnarray*}}

\newcommand{\bc}{\begin{center}}
\newcommand{\ec}{\end{center}}
\newcommand{\bi}{\begin{itemize}}
\newcommand{\ei}{\end{itemize}}
\newcommand{\benu}{\begin{enumerate}}
\newcommand{\eenu}{\end{enumerate}}
\newcommand{\bdes}{\begin{description}}
\newcommand{\edes}{\end{description}}
\newcommand{\bt}{\begin{tabular}}
\newcommand{\et}{\end{tabular}}

\newcommand \Sigmabf{\hbox{$\bf \Sigma$}}
\newcommand \Upsilonbf{\hbox{$\bf \Upsilon$}}

\newcommand \Deltabf{\hbox{$\bf \Delta$}}

\newcommand \Thetabf{\hbox{$\bf \Theta$}}
\newcommand \Lambdabf{\hbox{$\bf \Lambda$}}

\newcommand{\circlambda}{\mbox{$\Lambda$
             \kern-.85em\raise1.5ex
             \hbox{$\scriptstyle{\circ}$}}\,}

\renewcommand \Sigmabf{\boldsymbol{\Sigma}}
\renewcommand \Upsilonbf{\boldsymbol{\Upsilon}}

\renewcommand \Deltabf{\boldsymbol{\Delta}}

\renewcommand \Thetabf{\boldsymbol{\Theta}}
\renewcommand \Lambdabf{\boldsymbol{\Lambda}}

\usepackage{graphicx}
\usepackage{cite}
\usepackage[numbers,sort&compress]{natbib}

\begin{document}
\ArticleType{RESEARCH PAPER}
\Year{2024}
\Month{}
\Vol{}
\No{}
\DOI{}
\ArtNo{}
\ReceiveDate{}
\ReviseDate{}
\AcceptDate{}
\OnlineDate{}

\title{Bayesian Rao test for distributed target detection in interference and noise with limited training data}{Title keyword 5 for citation Title for citation Title for citation}

\author[1]{Daipeng Xiao}{}
\author[1]{Weijian Liu}{{liuvjian@163.com}}
\author[2]{Jun Liu}{}
\author[3]{Yuntao Wu}{}
\author[1]{Qinglei Du}{}
\author[4]{Xiaoqiang Hua}{}
\

\AuthorMark{}

\AuthorCitation{}


\address[1]{Wuhan Electronic Information Institute, Wuhan {\rm 430019}, China}
\address[2]{Department of Electronic Engineering and Information Science, University of Science and Technology of China,\\ Hefei {\rm 230027}, China}
\address[3]{School of Computer Science and Engineering, Wuhan Institute of Technology, Wuhan,  {\rm 430205}, China}
\address[4]{College of Electronic Science, National University of Defense Technology, Changsha {\rm 410073}, China}

\abstract{This paper has studied the problem of detecting a range-spread target in interference and noise when the number of training data is limited. The interference is located within a certain subspace with an unknown coordinate, while the noise follows a Gaussian distribution with an unknown covariance matrix. We concentrate on the scenarios where the training data are limited and employ a Bayesian framework to find a solution. Specifically, the covariance matrix is assumed to follow an inverse Wishart distribution. Then, we introduce the Bayesian detector according to the Rao test, which, demonstrated by both simulation experiment and real data, has superior detection performance to the existing detectors in certain situations.}


\maketitle

\section{Introduction}

Detecting signals in unknown noise is an essential issue in signal processing community \cite{LiuWu25SCIS,Jian23persymmetric,Liu2022,Long2019}. In 1986, Kelly proposed the well-known generalized likelihood ratio test (GLRT) for detecting signals in unknown Gaussian noise \cite{4104190}. Afterwards, an adaptive matched filter (AMF) was put forward \cite{135446} by building on the two-step GLRT. It was shown in \cite{1336827} that the AMF can also be derived from the Wald test. The Rao test corresponding to this scenario was provided in \cite{4244665}.

With advancements in radar technology, the performance of radar systems has been progressively enhanced, leading to a corresponding increase in radar resolution \cite{Liu2020a,Gao2016}. In high-resolution radar (HRR), it is widely known that a target is often extended in the range domain. Consequently, the target typically occupies several radar range resolution cells and is commonly referred to as a distributed target \cite{Jin2020,Jiang2023}.

It is worthy pointing out that there usually exists interference, intentionally or unintentionally.
In \cite{4133016}, to settle the matter of detection aimed at distributed target in subspace interference, several efficient detectors were proposed based on GLRT criterion. Furthermore, some detectors based on Rao test were given in \cite{LIU2015333}. Additionally, Liu \emph{et al.} investigated the detection problem in the presence of subspace interference and Gaussian noise in \cite{Liu2014a}, and proposed two types of adaptive detectors under the assumptions of known and unknown noise power, respectively. In \cite{Bandiera2013}, the authors assumed that the interference subspace was unknown and only the dimension of the subspace was known, and proposed a detector similar to the GLRT. In addition, it was assumed in \cite{Liu2016} that the interference was unknown but orthogonal to the signal in the whitened subspace, referred to as the orthogonal interference, which satisfied the generalized eigen-relation (GER) \cite{Richmond2000}. Then Liu \emph{et al.} proposed an adaptive detector based on the GLRT, whose detection performance was better than the corresponding Rao and Wald tests.

All of the detectors discussed above are proposed on the premise that there is an ample amount of training data, which is essential for reliably estimating the unknown covariance matrix, with the sample covariance matrix (SCM) being a well-known effective estimate. In particular, at least $N$ independent and identically distributed (IID) training data are needed to form an invertible SCM with probability one.
Nevertheless, in practice, this requirement may not always hold true due to the severe terrain undulations and the rapid changes in the spectral characteristics of clutter \cite{Xiao2024}. The actual environment may be sample-starved, a scenario where the number of training data is insufficient to reliably estimate the unknown covariance matrix. More specifically, in many real-world applications, the number of training data $L$ is less than the dimension of the test data $N$, i.e., $L < N$, which leads to the ineffectiveness of conventional detectors \cite{Liu2018}.

To handle the trouble of limited training data in signal detection, there are several approaches, one of which is the use of Bayesian theory. Within the framework of Bayesian theory \cite{Bandiera2015, Huang2022, Gao2017}, the unknown covariance matrix is considered as a random variable, which follows a certain statistical distribution. The parameters associated with this distribution are obtained from historical data or antenna configuration. Besson \emph{et al.} in \cite{4154721} introduced the first Bayesian detector for signal detection amidst unknown Gaussian noise. Other Bayesian detectors have been found in \cite{4359546, 5417154, 5371946, 9946376}.

In the available literature, relatively little attention has been given to the distributed target detection in presence of interference when training data are scarce. One study in \cite{Gao2017}  proposed two GLRT-based Bayesian detectors for this purpose. Due to the excessive number of unknown parameters in the detection process, there does not exist uniformly most powerful (UMP) test, it stands to reason to explore other criteria to develop novel detectors and evaluate their performance against the GLRT-based Bayesian detectors. Among such criteria, the Rao test is commonly used. Compared to the GLRT, the derivation of the detection statistic usually involves fewer unknown parameters to estimate. Hence, the Rao test is characterized by its lower computational complexity and relatively better stability in the presence of errors in practical environments. Accordingly, we propose a Bayesian detector, utilizing the Rao test, designed for the range-spread target detection amidst interference and noise environment where training data are limited. Experiments with simulated data and measured data both illustrate that the Bayesian Rao detector introduced in this paper exhibits superior detection performance compared to the GLRT-based Bayesian detectors previously given in \cite{Gao2017}. Additionally, our finding indicates that the Bayesian two-step GLRT (2S-GLRT) in \cite{Gao2017} can also be obtained through the application of the Wald test.

\section{Problem Formulation}
The distributed target detection problem in subspace can be expressed as the binary hypothesis test
\begin{equation}\label{1}
	\left\{ \begin{aligned}
		& {{\text{H}}_{1}}:\mathbf{Z}=\mathbf{\Phi{A}}+\mathbf{\Upsilon {W}}+\mathbf{N},~~{{\mathbf{Z}}_{L}}={{\mathbf{N}}_{L}} \\
		& {{\text{H}}_{0}}:\mathbf{Z}=\mathbf{\Upsilon {W}}+\mathbf{N},~~{{\mathbf{Z}}_{L}}={{\mathbf{N}}_{L}} \\
	\end{aligned} \right.\end{equation}
where ${{\text{H}}_{1}}$ denotes the hypothesis of signal presence and ${{\text{H}}_{0}}$ represents the hypothesis of signal absence, $\mathbf{Z}$ represents the test data matrix of $N\times K$ dimensions, $\mathbf{\Phi}$ is an $N\times p$ matrix with full-column-rank representing the signal, $\mathbf{\Upsilon }$ is an $N\times q$ matrix with full-column-rank representing the interference, the $p\times K$ matrix $\mathbf{A}$ and the $q\times K$ matrix $\mathbf{W}$ respectively symbolize the coordinates of the signal and the interference,  $\mathbf{N}$ represents the noise matrix in  the data under test, with each column of $\mathbf{N}$ being distributed in accordance with a complex Gaussian distribution with a mean of zero and an unspecified covariance matrix, denoted as $\mathbf{R}$. In order to get an estimate of $\mathbf{R}$, we need the training sample matrix ${{\mathbf{Z}}_{L}}$ which is of dimension $N\times L$, while ${{\mathbf{N}}_{L}}$ denotes the noise presented in training data. In \eqref{1}, $N$ represents the dimension of the data, $K$ represents the number of range cells occupied by the distributed target, $p$ is the dimension of the signal subspace, $q$ is the dimension of the interference subspace, and $L$ representing the amount of training data.

In many real complex and changing circumstances,  the limited size of training data often results in $L<N$. Consequently, the quantity of training data is insufficient to construct the inverse of sample covariance matrix (SCM). To resolve this issue, we adopt the Bayesian approach, where it is assumed that $\mathbf{R}$ is modeled as an inverse Wishart distribution, which has $\eta $ dimensions degrees of freedom (DOFs) and a scale matrix $\eta \Sigmabf $, it can be abbreviated as $\mathbf{R}\sim\mathcal{C}\mathcal{W}_{N}^{-1}\left( \eta ,\eta \Sigmabf  \right)$. Note that we have $\text{E}[\mathbf{R}]=\frac{\eta }{\eta -N}\Sigmabf $ and $\text{E}\left[ {{\left\| \mathbf{R}-\Sigmabf  \right\|}^{2}} \right]\approx \frac{1}{\eta }\text{t}{{\text{r}}^{2}}(\Sigmabf )$ \cite{5417154}, where $\text{E}[\cdot ]$ is statistical expectation, $||\cdot ||$ is the Frobenius norm, and $\text{tr}(\cdot )$ is the matrix trace. Essentially, a large $\eta $ means that the prior information of the statistical distribution for $\mathbf{R}$ is more reliable.

In practical applications, determining the degree of freedom $\eta$ can be based on prior knowledge or historical data. If historical data of the noise covariance matrix are available, we can analyze the distribution characteristics of the data and choose an appropriate $\eta$ value to make the inverse Wishart prior distribution better match the actual situation. Another approach is to use cross-validation methods. We can divide the available data into several subsets, use different $\eta$ values for training and testing on these subsets, and select the $\eta$ value that gives the best performance in terms of probability of detection (PD) or other evaluation metrics.

\section{The Proposed Bayesian Detector}
In this section, we devise an effective Bayesian detector designed for the problem formulated in \eqref{1} based on the Rao criterion. For convenience, we denote $\Thetabf $ as the parameter vector, having the form
\begin{equation}\label{2}
	\Thetabf ={{[\Thetabf _{\text{r}}^{T},\Thetabf _{\text{s}}^{T}]}^{T}}\end{equation}
with ${{\Thetabf }_{\text{r}}}=\text{vec}(\mathbf{A})$, ${{\Thetabf }_{\text{s}}}={{[\text{vec}^{T}(\mathbf{R}),\text{vec}^{T}(\mathbf{W})]}^{T}}$, $\text{vec}(\cdot )$ stands for vectorization operation, ${{(\cdot )}^{T}}$ denotes the transpose, ${{\Thetabf }_{\text{r}}}$ and ${{\Thetabf }_{\text{s}}}$ are respectively expressed as the correlated and uncorrelated parameters. It is acknowledged that the fisher information matrix (FIM) with respect to (w.r.t.) $\Thetabf $ is defined as \cite{Liu2014}
\begin{equation}\label{3}
	\mathbf{F}(\Thetabf )=\text{E}\left[ \frac{\partial \ln {{f}_{1}}(\mathbf{Z})}{\partial {{\Thetabf }^{*}}}\frac{\partial \ln {{f}_{1}}(\mathbf{Z})}{\partial {{\Thetabf }^{T}}} \right]\end{equation}
where $\partial (\cdot )$ is partial derivative, $\ln (\cdot )$ is natural logarithm, ${{(\cdot )}^{*}}$ is conjugate, and ${{(\cdot )}^{T}}$ is transpose. The FIM $\mathbf{F}(\Thetabf )$ in \eqref{3} is often partitioned in the following manner

\begin{equation}\label{4}
	\mathbf{F}(\Thetabf )=\left[ \begin{matrix}
		{{\mathbf{F}}_{{{\Thetabf }_{\text{r}}},{{\Thetabf }_{\text{r}}}}}(\Thetabf ) & {{\mathbf{F}}_{{{\Thetabf }_{\text{r}}},{{\Thetabf }_{\text{s}}}}}(\Thetabf )  \\
		{{\mathbf{F}}_{{{\Thetabf }_{\text{s}}},{{\Thetabf }_{\text{r}}}}}(\Thetabf ) & {{\mathbf{F}}_{{{\Thetabf }_{\text{s}}},{{\Thetabf }_{\text{s}}}}}(\Thetabf )  \\
	\end{matrix} \right]\end{equation}
We know that the Rao test can be indicated as
\begin{equation}\label{5}
	{{t}_{\text{Rao}}}=\left. \frac{\partial \ln {{f}_{1}}(\mathbf{Z})}{\partial {{\Thetabf }_{\text{r}}}} \right|_{\Thetabf ={{{\hat{\Thetabf }}}_{0}}}^{T}{{\left[ \mathbf{F}_{{}}^{-1}({{{\hat{\Thetabf }}}_{0}}) \right]}_{{{\Thetabf }_{\text{r}}},{{\Thetabf }_{\text{r}}}}}{{\left. \frac{\partial \ln {{f}_{1}}(\mathbf{Z})}{\partial \Thetabf _{\text{r}}^{*}} \right|}_{\Thetabf ={{{\hat{\Thetabf }}}_{0}}}}\end{equation}
where ${{\left[ \mathbf{F}_{{}}^{-1}({{{\hat{\Thetabf }}}_{0}}) \right]}_{{{\Thetabf }_{\text{r}}},{{\Thetabf }_{\text{r}}}}}$ is the value of
\begin{equation}\label{6}
	\begin{aligned}
		{{[\mathbf{F}^{-1}(\Thetabf )]}_{{{\Thetabf }_{\text{r}}},{{\Thetabf }_{\text{r}}}}}=&\left[ {{\mathbf{F}}_{{{\Thetabf }_{\text{r}}},{{\Thetabf }_{\text{r}}}}}(\Thetabf )-{{\mathbf{F}}_{{{\Thetabf }_{\text{r}}},{{\Thetabf }_{\text{s}}}}}(\Thetabf )\mathbf{F}_{{{\Thetabf }_{\text{s}}},{{\Thetabf }_{\text{s}}}}^{-1}
		(\Thetabf ){{\mathbf{F}}_{{{\Thetabf }_{\text{s}}},{{\Thetabf }_{\text{r}}}}}(\Thetabf ) \right]^{-1}
	\end{aligned}
\end{equation}
evaluated at ${{\hat{\Thetabf }}_{0}}$, i.e., a proper estimate of $\Thetabf $ under ${\text{H}_{0}}$.

For the problem indicated in \eqref{1}, we have the following probability density functions (PDFs)
\begin{equation}\label{7}
	{{f}_{0}}(\mathbf{Z}|\mathbf{R})=\frac{\text{etr}\left[ -\mathbf{R}_{{}}^{-1}(\mathbf{Z}-\mathbf{\Upsilon {W}}){{(\mathbf{Z}-\mathbf{\Upsilon {W}})}^{H}} \right]}{{{\pi }^{NK}}|\mathbf{R}{{|}^{K}}}\end{equation}
\begin{equation}\label{8}
	{{f}_{1}}(\mathbf{Z}|\mathbf{R})=\frac{\text{etr}\left[ -\mathbf{R}_{{}}^{-1}(\mathbf{Z}-\mathbf{BC}){{(\mathbf{Z}-\mathbf{BC})}^{H}} \right]}{{{\pi }^{NK}}|\mathbf{R}{{|}^{K}}}\end{equation}
\begin{equation}\label{9}
	f({{\mathbf{Z}}_{L}}|\mathbf{R})=\frac{\text{etr}(-\mathbf{R}_{{}}^{-1}\mathbf{S})}{{{\pi }^{NL}}|\mathbf{R}{{|}^{L}}}\end{equation}
\begin{equation}\label{10}
	f(\mathbf{R})=\frac{|\Sigmabf {{|}^{\eta }}}{c|\mathbf{R}{{|}^{\eta +N}}}\text{etr}\left( -\eta \Sigmabf \mathbf{R}_{{}}^{-1} \right)\end{equation}
where $\mathbf{B}=[\mathbf{\Phi},\mathbf{\Upsilon }]$, $\mathbf{C}={{[{{\mathbf{A}}^{T}},{{\mathbf{W}}^{T}}]}^{T}}$, $\mathbf{S}$ represents the SCM constructed from training data, $c$ is a constant, and $|\cdot |$ is the matrix determinant. It follows from \eqref{8} that
\begin{equation}\label{11}
	\frac{\partial \ln {{f}_{1}}(\mathbf{Z})}{\partial \text{vec}(\mathbf{A})}=\text{vec}\left[ {{(\mathbf{Z}_{1}^{H}\mathbf{R}_{{}}^{-1}\mathbf{\Phi})}^{T}} \right]\end{equation}
and
\begin{equation}\label{12}
	\frac{\partial \ln {{f}_{1}}(\mathbf{Z})}{\partial \text{vec}({{\mathbf{A}}^{*}})}=\text{vec}\left[ {{\mathbf{\Phi}}^{H}}\mathbf{R}_{{}}^{-1}{{\mathbf{Z}}_{1}} \right]\end{equation}
where ${{\mathbf{Z}}_{1}}=\mathbf{Z}-\mathbf{\Phi{A}}-\mathbf{\Upsilon {W}}$.
Taking a method in \cite{9591291} , we can obtain
\begin{equation}\label{13}
	\begin{aligned}
		{{[\mathbf{F}_{{}}^{-1}(\Thetabf )]}_{{{\Thetabf }_{\text{r}}},{{\Thetabf }_{\text{r}}}}}&={\left\{ {{\mathbf{I}}_{K}}\otimes \left[ {{\mathbf{\Phi}}^{H}}\mathbf{R}_{{}}^{-1}\mathbf{\Phi}-{{\mathbf{\Phi}}^{H}}\mathbf{R}_{{}}^{-1}\mathbf{\Upsilon }
			{{({{\mathbf{\Upsilon }}^{H}}\mathbf{R}_{{}}^{-1}\mathbf{\Upsilon })}^{-1}}{{\mathbf{\Upsilon }}^{H}}\mathbf{R}_{{}}^{-1}\mathbf{\Phi} \right] \right\}^{-1}} \\
		&={{\mathbf{I}}_{K}}\otimes {{({{\mathbf{\bar{\Phi}}}^{H}}\mathbf{P}_{{\mathbf{\bar{\Upsilon }}}}^{\bot }\mathbf{\bar{\Phi}})}^{-1}}
	\end{aligned}
\end{equation}
where $\mathbf{\bar{\Phi}}={{\mathbf{R}}^{-1/2}}\mathbf{\Phi}$, $\mathbf{\bar{\Upsilon }}={{\mathbf{R}}^{-1/2}}\mathbf{\Upsilon }$,$\mathbf{P}_{{\mathbf{\bar{\Upsilon }}}}^{\bot }={{\mathbf{I}}_{N}}-\mathbf{P}_{{\mathbf{\bar{\Upsilon }}}}^{{}}$, $\mathbf{P}_{{\mathbf{\bar{\Upsilon }}}}^{{}}=\mathbf{\bar{\Upsilon }}{{({{\mathbf{\bar{\Upsilon }}}^{H}}\mathbf{\bar{\Upsilon }})}^{-1}}{{\mathbf{\bar{\Upsilon }}}^{H}}$. By pluging \eqref{11}, \eqref{12} and \eqref{13} into \eqref{5} as well as letting $\mathbf{A}={\mathbf{0}_{p\times K}}$, we can derive the Rao test for the given $\mathbf{R}$ and $\mathbf{W}$ as follows
\begin{equation}\label{14}
\begin{aligned}
	{{t}_{\text{Ra}{{\text{o}}_{\mathbf{R,W}}}}} =&\text{ve}{{\text{c}}^{T}}\left[ {{\left( \mathbf{Z}_{0}^{H}{{\mathbf{R}}^{\mathbf{-1}}}\mathbf{\Phi} \right)}^{T}} \right]\left[ {{\mathbf{I}}_{K}}\otimes {{\left( {{{\mathbf{\bar{\Phi}}}}^{H}}\mathbf{P}_{{\mathbf{\bar{\Upsilon }}}}^{\bot }\mathbf{\bar{\Phi}} \right)}^{-1}} \right]
	\text{       }\text{vec}\left[ {{\mathbf{\Phi}}^{H}}{{\mathbf{R}}^{\text{-1}}}{{\mathbf{Z}}_{0}} \right] \\
	=&\text{ve}{{\text{c}}^{T}}\left[ {{\left( \mathbf{\bar{Z}}_{0}^{H}\mathbf{\bar{\Phi}} \right)}^{T}} \right]\text{vec}\left[ {{\left( {{{\mathbf{\bar{\Phi}}}}^{H}}\mathbf{P}_{{\mathbf{\bar{\Upsilon }}}}^{\bot }\mathbf{\bar{\Phi}} \right)}^{-1}}{{{\mathbf{\bar{\Phi}}}}^{H}}{{{\mathbf{\bar{Z}}}}_{0}} \right] \\
	=&\text{tr}\left[ \mathbf{\bar{Z}}_{0}^{H}\mathbf{\bar{\Phi}}{{\left( {{{\mathbf{\bar{\Phi}}}}^{H}}\mathbf{P}_{{\mathbf{\bar{\Upsilon }}}}^{\bot }\mathbf{\bar{\Phi}} \right)}^{-1}}{{{\mathbf{\bar{\Phi}}}}^{H}}{{{\mathbf{\bar{Z}}}}_{0}} \right]  	
\end{aligned}\end{equation}
where ${{\mathbf{Z}}_{0}}=\mathbf{Z}-\mathbf{\Upsilon {W}}$ and we have used $\text{vec}({{\mathbf{D}}_{1}}{{\mathbf{D}}_{2}}{{\mathbf{D}}_{3}})=(\mathbf{D}_{3}^{T}\otimes
{{\mathbf{D}}_{1}})\text{vec}({{\mathbf{D}}_{2}})$ and $\text{ve}{{\text{c}}^{T}}(\mathbf{D}_{4}^{T})\text{vec}({{\mathbf{D}}_{5}})=\text{tr}({{\mathbf{D}}_{4}}{{\mathbf{D}}_{5}})$ for any comparable matrix.

According to \eqref{7}, \eqref{9} and \eqref{10}, we have
\begin{equation}\label{15}
\begin{aligned}
	{{f}_{0}}\left( \mathbf{Z}|\mathbf{R} \right)f({{\mathbf{Z}}_{L}}|\mathbf{R})f(\mathbf{R})& =c\frac{{{\left| \mathbf{\Sigma } \right|}^{\eta}}\text{etr}\left[ -{{\mathbf{R}}^{-1}}\left( {{\mathbf{Z}}_{0}}\mathbf{Z}_{0}^{H}+\mathbf{S}+\eta \mathbf{\Sigma } \right) \right]}{{{\left| \mathbf{R} \right|}^{\eta +N+L+K}}}
	& \text{               }
\end{aligned}
\end{equation}
Nulling the partial derivative of the logarithm of \eqref{15} w.r.t. $\mathbf{W}$, we have the maximum likelihood estimate (MLE) of $\mathbf{W}$ for given $\mathbf{R}$ as follows
\begin{equation}\label{16}
\mathbf{\hat{W}}={{({{\mathbf{\Upsilon }}^{H}}{{\mathbf{R}}^{-1}}\mathbf{\Upsilon })}^{-1}}{{\mathbf{\Upsilon }}^{H}} {{\mathbf{R}}^{-1}}\mathbf{Z}={{({{\mathbf{\bar{\Upsilon }}}^{H}}\mathbf{\bar{\Upsilon }})}^{-1}} {{\mathbf{\bar{\Upsilon }}}^{H}}\mathbf{\bar{Z}}\end{equation}
By plugging \eqref{16} into \eqref{14}, we get the Rao test through calculation for given $\mathbf{R}$ as
\begin{equation}\label{17}
{{t}_{{\text{Rao}_{\mathbf{R}}}}}=\text{tr}\left[ \mathbf{\bar{Z}}_{{}}^{H}\mathbf{P}_{{\mathbf{\bar{\Upsilon }}}}^{\bot }\mathbf{\bar{\Phi}}{{({{{\mathbf{\bar{\Phi}}}}^{H}}\mathbf{P}_{{\mathbf{\bar{\Upsilon }}}}^{\bot }\mathbf{\bar{\Phi}})}^{-1}}{{{\mathbf{\bar{\Phi}}}}^{H}}\mathbf{P}_{{\mathbf{\bar{\Upsilon }}}}^{\bot }\mathbf{\bar{Z}} \right]\end{equation}
For convenience, we express \eqref{17} as
\begin{equation}\label{18}
{t}_{{\text{Rao}}_{\mathbf{R}}}=\text{tr}({{\Deltabf }^{H}}\Lambdabf \Deltabf )\end{equation}
where
\begin{equation}\label{19}
\Deltabf ={{\mathbf{\Phi}}^{H}}{{\mathbf{R}}^{-1}}\left[ \mathbf{Z}-\mathbf{\Upsilon }{{({{\mathbf{\Upsilon }}^{H}}{{\mathbf{R}}^{-1}}\mathbf{\Upsilon })}^{-1}}{{\mathbf{R}}^{-1}}\mathbf{Z} \right]\end{equation}
and
\begin{equation}\label{20}
\Lambdabf ={{\left[ {{\mathbf{\Phi}}^{H}}{{\mathbf{R}}^{-1}}\mathbf{\Phi}-{{\mathbf{\Phi}}^{H}}{{\mathbf{R}}^{-1}}\mathbf{\Upsilon } {{({{\mathbf{\Upsilon }}^{H}}{{\mathbf{R}}^{-1}}\mathbf{\Upsilon })}^{-1}}\Upsilonbf^H{{\mathbf{R}}^{-1}}\mathbf{\Phi} \right]}^{-1}}\end{equation}
To get the final Bayesian Rao test, an estimate of $\mathbf{R}$ is required under ${{\text{H}}_{0}}$. Letting the derive of \eqref{15} w.r.t. $\mathbf{R}$ to be zero under ${{\text{H}}_{0}}$, we can get the maximum a posteriori (MAP) estimate of $\mathbf{R}$ under ${{\text{H}}_{0}}$ for given $\mathbf{W}$ as
\begin{equation}\label{21}
\begin{aligned}
	{{{\mathbf{\hat{R}}}}_{0}} &=\frac{(\mathbf{Z}-\mathbf{\Upsilon {W}}){{(\mathbf{Z}-\mathbf{\Upsilon {W}})}^{H}}+\mathbf{S}+\eta\Sigmabf }{\eta +N+L+K} \\
	& ={{(\mathbf{S}+\eta \Sigmabf )}^{1/2}}\frac{(\mathbf{\breve{Z}}-\mathbf{\breve{\Upsilon }W}){{(\mathbf{\breve{Z}}-\mathbf{\breve{\Upsilon }W})}^{H}}+{{\mathbf{I}}_{N}}}{\eta +N+L+K}{{(\mathbf{S}+\eta \Sigmabf )}^{1/2}}
\end{aligned}\end{equation}
We proceed to derive the MAP of $\mathbf{W}$. Performing the integration of \eqref{15} w.r.t. $\mathbf{R}$ leads to
\begin{equation}\label{22}
\begin{aligned}
	& \int{{{f}_{0}}(\mathbf{Z}|\mathbf{R})f({{\mathbf{Z}}_{L}}|\mathbf{R})f(\mathbf{R})\text{d}\mathbf{R}} \\
	& =c\frac{{{\left| \left. \Sigmabf  \right| \right.}^{\eta }}}{|{{\mathbf{Z}}_{0}}\mathbf{Z}_{0}^{H}+\mathbf{S}+\eta \Sigmabf {{|}^{\eta +L+K}}}\int{\frac{|{{\mathbf{Z}}_{0}}\mathbf{Z}_{0}^{H}+\mathbf{S}+\eta \Sigmabf {{|}^{\eta +L+K}}}{|\mathbf{R}{{|}^{\eta +N+L+K}}}} \\
	& \times \text{ etr}\left[ -{{\mathbf{R}}^{-1}}\left( {{\mathbf{Z}}_{0}}\mathbf{Z}_{0}^{H}+\mathbf{S}+\eta \Sigmabf  \right) \right]\text{d}\mathbf{R} \\
	& =c\frac{{{\left| \left. \Sigmabf  \right| \right.}^{\eta }}}{|{{\mathbf{Z}}_{0}}\mathbf{Z}_{0}^{H}+\mathbf{S}+\eta \Sigmabf {{|}^{\eta +L+K}}} \\
\end{aligned}\end{equation}
One can verify that
\begin{equation}\label{23}
\begin{aligned}
	\left| {{\mathbf{Z}}_{0}}\mathbf{Z}_{0}^{H}+\mathbf{S}+\eta \Sigmabf  \right| &=\left| (\mathbf{Z}-\mathbf{\Upsilon {W}}){{(\mathbf{Z}-\mathbf{\Upsilon {W}})}^{H}}+\mathbf{S}+\eta \Sigmabf  \right| \\
	& =\left| \mathbf{S}+\eta \Sigmabf  \right|\cdot \left| {{\mathbf{I}}_{K}}+{{(\mathbf{Z}-\mathbf{\Upsilon {W})}}^{H}}{{(\mathbf{S}+\eta \Sigmabf )}^{-1}}(\mathbf{Z}-\mathbf{\Upsilon {W}}) \right| \\
	& =\left| \mathbf{S}+\eta \Sigmabf  \right|\cdot \left| {{\mathbf{I}}_{K}}+{{(\mathbf{\breve{Z}}-\mathbf{\breve{\Upsilon }W})}^{H}}(\mathbf{\breve{Z}}-\mathbf{\breve{\Upsilon }W}) \right|
\end{aligned}\end{equation}
where $\mathbf{\breve{Z}}={{(\mathbf{S}+\eta \Sigmabf )}^{-1/2}}\mathbf{Z}$, $\mathbf{\breve{\Upsilon }}={{(\mathbf{S}+\eta \Sigmabf )}^{-1/2}}\mathbf{\Upsilon }$, and $\mathbf{\breve{\Phi}}={{(\mathbf{S}+\eta \Sigmabf )}^{-1/2}}\mathbf{\Phi}$. Letting the derivative of \eqref{23} w.r.t. $\mathbf{W}$ be zero results in the MAP of $\mathbf{W}$ as
\begin{equation}\label{24}
\mathbf{\hat{W}}={{({{\mathbf{\breve{\Upsilon }}}^{H}}\mathbf{\breve{\Upsilon }})}^{-1}}{{\mathbf{\breve{\Upsilon }}}^{H}}\mathbf{\breve{Z}}\end{equation}
Substituting \eqref{24} into \eqref{21} yields the MAP of $\mathbf{R}$ under ${{\text{H}}_{0}}$ as
\begin{equation}\label{25}
{{\mathbf{\hat{R}}}_{0}}={{(\mathbf{S}+\eta \Sigmabf )}^{1/2}}\frac{\mathbf{P}_{{\mathbf{\breve{\Upsilon }}}}^{\bot }\mathbf{\breve{Z}}{{{\mathbf{\breve{Z}}}}^{H}}\mathbf{P}_{{\mathbf{\breve{\Upsilon }}}}^{\bot }+{{\mathbf{I}}_{N}}}{\eta +N+L+K}{{(\mathbf{S}+\eta \Sigmabf )}^{1/2}}\end{equation}
where $\mathbf{P}_{{\mathbf{\breve{\Upsilon }}}}^{\bot }={{\mathbf{I}}_{N}}-\mathbf{P}_{{\mathbf{\breve{\Upsilon }}}}^{{}}$ and $\mathbf{P}_{{\mathbf{\breve{\Upsilon }}}}^{{}}=\mathbf{\breve{\Upsilon }}{{({{\mathbf{\breve{\Upsilon }}}^{H}}\mathbf{\breve{\Upsilon }})}^{-1}}{{\mathbf{\breve{\Upsilon }}}^{H}}$. Performing the matrix inversion to \eqref{25} yields
\begin{equation}\label{26}
\begin{aligned}
	\mathbf{\hat{R}}_{0}^{-1}=&\alpha {{(\mathbf{S}+\eta \Sigmabf )}^{-1/2}}\left[ {{\mathbf{I}}_{N}}-\mathbf{P}_{{\mathbf{\breve{\Upsilon }}}}^{\bot }\mathbf{\breve{Z}}{{({{\mathbf{I}}_{K}}+{{{\mathbf{\breve{Z}}}}^{H}}\mathbf{P}_{{\mathbf{\breve{\Upsilon }}}}^{\bot }\mathbf{\breve{Z}})}^{-1}}
	{{{\mathbf{\breve{Z}}}}^{H}}\mathbf{P}_{{\mathbf{\breve{\Upsilon }}}}^{\bot } \right]{{(\mathbf{S}+\eta \Sigmabf )}^{-1/2}}
\end{aligned}
\end{equation}
where $\alpha ={{(\eta +N+L+K)}}$. It follows from \eqref{26} that
\begin{equation}\label{27}
\mathbf{\hat{R}}_{0}^{-1}\mathbf{\Upsilon }=\alpha {{(\mathbf{S}+\eta \Sigmabf )}^{-1}}\mathbf{\Upsilon }\end{equation}
Using \eqref{26} and \eqref{27}, we can rewrite \eqref{19} and \eqref{20} as\begin{equation}\label{28}
\begin{aligned}
	\Deltabf &=\alpha {{{\mathbf{\breve{\Phi}}}}^{H}}\left[ {{\mathbf{I}}_{N}}-\mathbf{P}_{{\mathbf{\breve{\Upsilon }}}}^{\bot }\mathbf{\breve{Z}}{{({{\mathbf{I}}_{K}}+{{{\mathbf{\breve{Z}}}}^{H}}\mathbf{P}_{{\mathbf{\breve{\Upsilon }}}}^{\bot }\mathbf{\breve{Z}})}^{-1}}{{{\mathbf{\breve{Z}}}}^{H}}\mathbf{P}_{{\mathbf{\breve{\Upsilon }}}}^{\bot } \right]\mathbf{P}_{{\mathbf{\breve{\Upsilon }}}}^{\bot }\mathbf{\breve{Z}} \\
	& =\alpha \left[ {{{\mathbf{\breve{\Phi}}}}^{H}}\mathbf{P}_{{\mathbf{\breve{\Upsilon }}}}^{\bot }\mathbf{\breve{Z}}-{{{\mathbf{\breve{\Phi}}}}^{H}}\mathbf{P}_{{\mathbf{\breve{\Upsilon }}}}^{\bot }\mathbf{\breve{Z}}{{({{\mathbf{I}}_{K}}+{{{\mathbf{\breve{Z}}}}^{H}}\mathbf{P}_{{\mathbf{\breve{\Upsilon }}}}^{\bot }\mathbf{\breve{Z}})}^{-1}}{{{\mathbf{\breve{Z}}}}^{H}}\mathbf{P}_{{\mathbf{\breve{\Upsilon }}}}^{\bot }\mathbf{\breve{Z}} \right] \\
	& =\alpha {{{\mathbf{\breve{\Phi}}}}^{H}}\mathbf{P}_{{\mathbf{\breve{\Upsilon }}}}^{\bot }\mathbf{\breve{Z}}{{({{\mathbf{I}}_{K}}+{{{\mathbf{\breve{Z}}}}^{H}}\mathbf{P}_{{\mathbf{\breve{\Upsilon }}}}^{\bot }\mathbf{\breve{Z}})}^{-1}}
\end{aligned}
\end{equation}
and
\begin{equation}\label{29}
\begin{aligned}
	\Lambdabf =
	&\alpha \left\{ {{{\mathbf{\breve{\Phi}}}}^{H}}\left[ {{\mathbf{I}}_{N}}-\mathbf{P}_{{\mathbf{\breve{\Upsilon }}}}^{\bot }\mathbf{\breve{Z}}{{({{\mathbf{I}}_{K}}+{{{\mathbf{\breve{Z}}}}^{H}}\mathbf{P}_{{\mathbf{\breve{\Upsilon }}}}^{\bot }\mathbf{\breve{Z}})}^{-1}}{{{\mathbf{\breve{Z}}}}^{H}}\mathbf{P}_{{\mathbf{\breve{\Upsilon }}}}^{\bot } \right]\mathbf{\breve{\Phi}} \right. \\
	& -{{{\mathbf{\breve{\Phi}}}}^{H}}\mathbf{P}_{{\mathbf{\breve{\Upsilon }}}}^{{}}{{\left. \left[ {{\mathbf{I}}_{N}}-\mathbf{P}_{{\mathbf{\breve{\Upsilon }}}}^{\bot }\mathbf{\breve{Z}}{{({{\mathbf{I}}_{K}}+{{{\mathbf{\breve{Z}}}}^{H}} \mathbf{P}_{{\mathbf{\breve{\Upsilon }}}}^{\bot }\mathbf{\breve{Z}})}^{-1}}{{{\mathbf{\breve{Z}}}}^{H}}\mathbf{P}_{{\mathbf{\breve{\Upsilon }}}}^{\bot } \right]\mathbf{\breve{\Phi}} \right\}}^{-1}} \\
	=&\alpha \left\{ {{{\mathbf{\breve{\Phi}}}}^{H}}\left[ {{\mathbf{I}}_{N}}-\mathbf{P}_{{\mathbf{\breve{\Upsilon }}}}^{\bot }\mathbf{\breve{Z}}{{({{\mathbf{I}}_{K}}+{{{\mathbf{\breve{Z}}}}^{H}}\mathbf{P}_{{\mathbf{\breve{\Upsilon }}}}^{\bot }\mathbf{\breve{Z}})}^{-1}}{{{\mathbf{\breve{Z}}}}^{H}}\mathbf{P}_{{\mathbf{\breve{\Upsilon }}}}^{\bot } \right]\mathbf{\breve{\Phi}} \right.-{{{\mathbf{\breve{\Phi}}}}^{H}}\mathbf{P}_{{\mathbf{\breve{\Upsilon }}}}^{{}}{{\left. {\mathbf{\breve{\Phi}}} \right\}}^{-1}} \\
	=&\alpha {{\left[ {{{\mathbf{\breve{\Phi}}}}^{H}}\mathbf{P}_{{\mathbf{\breve{\Upsilon }}}}^{\bot }\mathbf{\breve{\Phi}}-{{{\mathbf{\breve{\Phi}}}}^{H}}\mathbf{P}_{{\mathbf{\breve{\Upsilon }}}}^{\bot }\mathbf{\breve{Z}}{{({{\mathbf{I}}_{K}}+{{{\mathbf{\breve{Z}}}}^{H}}\mathbf{P}_{{\mathbf{\breve{\Upsilon }}}}^{\bot }\mathbf{\breve{Z}})}^{-1}}{{{\mathbf{\breve{Z}}}}^{H}}\mathbf{P}_{{\mathbf{\breve{\Upsilon }}}}^{\bot }\mathbf{\breve{\Phi}} \right]}^{-1}} \\
\end{aligned}
\end{equation}
respectively, where ${{\mathbf{P}}_{\mathbf{P}_{{\mathbf{\breve{\Upsilon }}}}^{\bot }\mathbf{\breve{\Phi}}}}=\mathbf{P}_{{\mathbf{\breve{\Upsilon }}}}^{\bot }\mathbf{\breve{\Phi}}{{({{\mathbf{\breve{\Phi}}}^{H}}\mathbf{P}_{{\mathbf{\breve{\Upsilon }}}}^{\bot }\mathbf{\breve{\Phi}})}^{-1}}{{\mathbf{\breve{\Phi}}}^{H}}\mathbf{P}_{{\mathbf{\breve{\Upsilon }}}}^{\bot }$, and in \eqref{28} we utilized equation ${{({{\mathbf{I}}_{K}}+{{\mathbf{\breve{Z}}}^{H}}\mathbf{P}_{{\mathbf{\breve{\Upsilon }}}}^{\bot }\mathbf{\breve{Z}})}^{-1}}={{\mathbf{I}}_{K}}- {{({{\mathbf{I}}_{K}}+{{\mathbf{\breve{Z}}}^{H}}\mathbf{P}_{{\mathbf{\breve{\Upsilon }}}}^{\bot }\mathbf{\breve{Z}})}^{-1}}{{\mathbf{\breve{Z}}}^{H}}\mathbf{P}_{{\mathbf{\breve{\Upsilon }}}}^{\bot }\mathbf{\breve{Z}}$.

Using matrix inversion lemma, i.e., the Sherman-Morrison-Woodbury formula, we can rewrite \eqref{29} as
\begin{equation}\label{29x}
\begin{aligned}
	\Lambdabf  
	=&\alpha \left\{ {{({{{\mathbf{\breve{\Phi}}}}^{H}}\mathbf{P}_{{\mathbf{\breve{\Upsilon }}}}^{\bot }\mathbf{\breve{\Phi}})}^{-1}}+{{({{{\mathbf{\breve{\Phi}}}}^{H}}\mathbf{P}_{{\mathbf{\breve{\Upsilon }}}}^{\bot }\mathbf{\breve{\Phi}})}^{-1}}{{{\mathbf{\breve{\Phi}}}}^{H}}\mathbf{P}_{{\mathbf{\breve{\Upsilon }}}}^{\bot }\mathbf{\breve{Z}}\right.\\
	&\left.     \cdot\text{               } {{({{\mathbf{I}}_{K}}+{{{\mathbf{\breve{Z}}}}^{H}}\mathbf{P}_{{\mathbf{\breve{\Upsilon }}}}^{\bot }\mathbf{\breve{Z}}-{{{\mathbf{\breve{Z}}}}^{H}}{{\mathbf{P}}_{\mathbf{P}_{{\mathbf{\breve{\Upsilon }}}}^{\bot }\mathbf{\breve{\Phi}}}}\mathbf{\breve{Z}})}^{-1}}{{{\mathbf{\breve{Z}}}}^{H}}\mathbf{P}_{{\mathbf{\breve{\Upsilon }}}}^{\bot }\mathbf{\breve{\Phi}}{{({{{\mathbf{\breve{\Phi}}}}^{H}}\mathbf{P}_{{\mathbf{\breve{\Upsilon }}}}^{\bot }\mathbf{\breve{\Phi}})}^{-1}} \right\}
\end{aligned}
\end{equation}

Substituting \eqref{28} and \eqref{29x} into \eqref{18} and ignoring the constant leads to
\begin{equation}\label{30}
\begin{aligned}
		& {{t}_{{\text{B-Rao-I}}}}
		  =\text{tr}\left\{ {{({{\mathbf{I}}_{K}}+{{{\mathbf{\breve{Z}}}}^{H}}\mathbf{P}_{{\mathbf{\breve{\Upsilon }}}}^{\bot }\mathbf{\breve{Z}})}^{-1}}{{{\mathbf{\breve{Z}}}}^{H}} \right. \\
		&\cdot \left.\left[ {{\mathbf{P}}_{\mathbf{P}_{{\mathbf{\breve{\Upsilon }}}}^{\bot }\mathbf{\breve{\Phi}}}}+{{\mathbf{P}}_{\mathbf{P}_{{\mathbf{\breve{\Upsilon }}}}^{\bot }\mathbf{\breve{\Phi}}}}\mathbf{\breve{Z}}{{({{\mathbf{I}}_{K}}+{{{\mathbf{\breve{Z}}}}^{H}}\mathbf{P}_{{\mathbf{\breve{\Upsilon }}}}^{\bot }\mathbf{\breve{Z}}-{{{\mathbf{\breve{Z}}}}^{H}}{{\mathbf{P}}_{\mathbf{P}_{{\mathbf{\breve{\Upsilon }}}}^{\bot }\mathbf{\breve{\Phi}}}}\mathbf{\breve{Z}})}^{-1}}{{{\mathbf{\breve{Z}}}}^{H}}{{\mathbf{P}}_{\mathbf{P}_{{\mathbf{\breve{\Upsilon }}}}^{\bot }\mathbf{\breve{\Phi}}}} \right]
		\mathbf{\breve{Z}}{{({{\mathbf{I}}_{K}}+{{{\mathbf{\breve{Z}}}}^{H}}\mathbf{P}_{{\mathbf{\breve{\Upsilon }}}}^{\bot }\mathbf{\breve{Z}})}^{-1}} \right\} \\
\end{aligned}
\end{equation}
Using the property of matrix trace, we can express \eqref{30} as
\begin{equation}\label{30x}
\begin{aligned}
		 {{t}_{{\text{B-Rao-I}}}} 
		& =\text{tr}\left\{ {{({{\mathbf{I}}_{K}}+{{{\mathbf{\breve{Z}}}}^{H}}\mathbf{P}_{{\mathbf{\breve{\Upsilon }}}}^{\bot }\mathbf{\breve{Z}})}^{-1}}{{{\mathbf{\breve{Z}}}}^{H}}{{\mathbf{P}}_{\mathbf{P}_{{\mathbf{\breve{\Upsilon }}}}^{\bot }\mathbf{\breve{\Phi}}}}\mathbf{\breve{Z}} \right. \\
		&\cdot \left.\left[ {{\mathbf{I}}_{K}}+{{({{\mathbf{I}}_{K}}+{{{\mathbf{\breve{Z}}}}^{H}}\mathbf{P}_{{\mathbf{\breve{\Upsilon }}}}^{\bot }\mathbf{\breve{Z}}-{{{\mathbf{\breve{Z}}}}^{H}}{{\mathbf{P}}_{\mathbf{P}_{{\mathbf{\breve{\Upsilon }}}}^{\bot }\mathbf{\breve{\Phi}}}}\mathbf{\breve{Z}})}^{-1}}{{{\mathbf{\breve{Z}}}}^{H}}{{\mathbf{P}}_{\mathbf{P}_{{\mathbf{\breve{\Upsilon }}}}^{\bot }\mathbf{\breve{\Phi}}}}\mathbf{\breve{Z}} \right]  {{({{\mathbf{I}}_{K}}+{{{\mathbf{\breve{Z}}}}^{H}}\mathbf{P}_{{\mathbf{\breve{\Upsilon }}}}^{\bot }\mathbf{\breve{Z}})}^{-1}} \right\}
\end{aligned}
\end{equation}
which is referred to as the Bayesian Rao test with interference rejection (B-Rao-I).
One can verify that
\begin{equation}\label{31}
\begin{aligned}
	& {{\mathbf{I}}_{K}}+{{({{\mathbf{I}}_{K}}+{{{\mathbf{\breve{Z}}}}^{H}}\mathbf{P}_{{\mathbf{\breve{\Upsilon }}}}^{\bot }\mathbf{\breve{Z}}-{{{\mathbf{\breve{Z}}}}^{H}}{{\mathbf{P}}_{\mathbf{P}_{{\mathbf{\breve{\Upsilon }}}}^{\bot }\mathbf{\breve{\Phi}}}}\mathbf{\breve{Z}})}^{-1}}{{{\mathbf{\breve{Z}}}}^{H}}{{\mathbf{P}}_{\mathbf{P}_{{\mathbf{\breve{\Upsilon }}}}^{\bot }\mathbf{\breve{\Phi}}}}\mathbf{\breve{Z}} \\
	& ={{({{\mathbf{I}}_{K}}+{{{\mathbf{\breve{Z}}}}^{H}}\mathbf{P}_{{\mathbf{\breve{\Upsilon }}}}^{\bot }\mathbf{\breve{Z}}- {{{\mathbf{\breve{Z}}}}^{H}}{{\mathbf{P}}_{\mathbf{P}_{{\mathbf{\breve{\Upsilon }}}}^{\bot }\mathbf{\breve{\Phi}}}}\mathbf{\breve{Z}})}^{-1}} ({{\mathbf{I}}_{K}}+{{{\mathbf{\breve{Z}}}}^{H}}\mathbf{P}_{{\mathbf{\breve{\Upsilon }}}}^{\bot }\mathbf{\breve{Z}}-{{{\mathbf{\breve{Z}}}}^{H}}{{\mathbf{P}}_{\mathbf{P}_{{\mathbf{\breve{\Upsilon }}}}^{\bot }\mathbf{\breve{\Phi}}}}\mathbf{\breve{Z}}+{{{\mathbf{\breve{Z}}}}^{H}}{{\mathbf{P}}_{\mathbf{P}_{{\mathbf{\breve{\Upsilon }}}}^{\bot }\mathbf{\breve{\Phi}}}}\mathbf{\breve{Z}}) \\
	&= {{({{\mathbf{I}}_{K}}+{{{\mathbf{\breve{Z}}}}^{H}}\mathbf{P}_{{\mathbf{\breve{\Upsilon }}}}^{\bot }\mathbf{\breve{Z}}-{{{\mathbf{\breve{Z}}}}^{H}}{{\mathbf{P}}_{\mathbf{P}_{{\mathbf{\breve{\Upsilon }}}}^{\bot }\mathbf{\breve{\Phi}}}}\mathbf{\breve{Z}})}^{-1}}({{\mathbf{I}}_{K}}+{{{\mathbf{\breve{Z}}}}^{H}}\mathbf{P}_{{\mathbf{\breve{\Upsilon }}}}^{\bot }\mathbf{\breve{Z}}) \\
\end{aligned}\end{equation}
Substituting \eqref{31} into \eqref{30} results in the compact form of the B-Rao-I
\begin{equation}\label{32}
\begin{aligned}
	{{t}_{\text{B-Rao-I}}}=\text{tr}\left[ {{({{\mathbf{I}}_{K}}+{{{\mathbf{\breve{Z}}}}^{H}}\mathbf{P}_{{\mathbf{\breve{\Upsilon }}}}^{\bot }\mathbf{\breve{Z}})}^{-1}}{{{\mathbf{\breve{Z}}}}^{H}}{{\mathbf{P}}_{\mathbf{P}_{{\mathbf{\breve{\Upsilon }}}}^{\bot }\mathbf{\breve{\Phi}}}}\mathbf{\breve{Z}}
	{({{\mathbf{I}}_{K}}+{{{\mathbf{\breve{Z}}}}^{H}}\mathbf{P}_{{\mathbf{\breve{\Upsilon }}}}^{\bot }\mathbf{\breve{Z}}-{{{\mathbf{\breve{Z}}}}^{H}}{{\mathbf{P}}_{\mathbf{P}_{{\mathbf{\breve{\Upsilon }}}}^{\bot }\mathbf{\breve{\Phi}}}}\mathbf{\breve{Z}})}^{-1} \right]
\end{aligned}
\end{equation}
Moreover, according to the identity $\mathbf{P}_{{\mathbf{\breve{B}}}}^{\bot }=\mathbf{P}_{{\mathbf{\breve{\Upsilon }}}}^{\bot }-{{\mathbf{P}}_{\mathbf{P}_{{\mathbf{\breve{\Upsilon }}}}^{\bot }\mathbf{\breve{\Phi}}}}$, we can rewrite \eqref{32} as
\begin{equation}\label{33}
\begin{aligned}
	{{t}_{\text{B-Rao-I}}}=&\text{tr}\left[ {{({{\mathbf{I}}_{K}}+{{{\mathbf{\breve{Z}}}}^{H}}\mathbf{P}_{{\mathbf{\breve{\Upsilon }}}}^{\bot }\mathbf{\breve{Z}})}^{-1}}{{{\mathbf{\breve{Z}}}}^{H}}{{\mathbf{P}}_{\mathbf{P}_{{\mathbf{\breve{\Upsilon }}}}^{\bot }\mathbf{\breve{\Phi}}}}\mathbf{\breve{Z}}
	{({{\mathbf{I}}_{K}}+{{{\mathbf{\breve{Z}}}}^{H}}\mathbf{P}_{{\mathbf{\breve{B}}}}^{\bot }\mathbf{\breve{Z}})}^{-1} \right]
\end{aligned}
\end{equation}

Note that the B-Rao-I is derived under the assumption of a homogeneous environment. In cases where the environment is heterogeneous, we can utilize robust estimation techniques and compensation methods among others. Specifically, robust estimation techniques are shown to be effective in addressing heterogeneous clutter, examples of which include the normalized sample covariance matrix (NSCM) \cite{ConteLops96} and the fixed-point estimator (FPE) \cite{GiniGreco02a}. These methods leverage heterogeneous data to mitigate the impacts of heterogeneity. Alternatively, the compensation method models the mismatch in the covariance matrix, estimates it using suitable techniques, and subsequently compensates for it, thereby enabling an effective estimation of the unknown covariance matrix using heterogeneous data \cite{Richmond00a,Besson07b}.

\section{Performance Evaluation}
In this section, we assess the performance of B-Rao-I using both simulated and actual data. To facilitate comparison, we take into account the ordinary GLRT and 2S-GLRT proposed in regard to the detection problem in \eqref{1}, described as \cite{4133016}
\begin{equation}\label{34}
	{{t}_{\text{GLRT-I}}}=\frac{\left| {{\mathbf{I}}_{K}}+{{{\mathbf{\tilde{Z}}}}^{H}}\mathbf{P}_{{\mathbf{\tilde{\Upsilon }}}}^{\bot }\mathbf{\tilde{Z}} \right|}{\left| {{\mathbf{I}}_{K}}+{{{\mathbf{\tilde{Z}}}}^{H}}\mathbf{P}_{{\mathbf{\tilde{B}}}}^{\bot }\mathbf{\tilde{Z}} \right|}\end{equation}
and
\begin{equation}\label{35}
	{{t}_\text{2S-GLRT-I}}=\text{tr}\left[ {{{\mathbf{\tilde{Z}}}}^{H}}\mathbf{P}_{{\mathbf{\tilde{\Upsilon }}}}^{\bot }\mathbf{\tilde{\Phi}}{{({{{\mathbf{\tilde{\Phi}}}}^{H}}\mathbf{P}_{{\mathbf{\tilde{\Upsilon }}}}^{\bot }\mathbf{\tilde{\Phi}})}^{-1}}{{{\mathbf{\tilde{\Phi}}}}^{H}}\mathbf{P}_{{\mathbf{\tilde{\Upsilon }}}}^{\bot }\mathbf{\tilde{Z}} \right]\end{equation}
respectively, where $\mathbf{\tilde{Z}}={{\mathbf{S}}^{-1/2}}\mathbf{Z}$, $\mathbf{\tilde{\Phi}}={{\mathbf{S}}^{-1/2}}\mathbf{\Phi}$, $\mathbf{\tilde{\Upsilon
}}={{\mathbf{S}}^{-1/2}}\mathbf{\Upsilon }$, $\mathbf{\tilde{B}}={{\mathbf{S}}^{-1/2}}\mathbf{B}$, $\mathbf{P}_{{\mathbf{\tilde{\Upsilon }}}}^{\bot }={{\mathbf{I}}_{N}}-\mathbf{P}_{{\mathbf{\tilde{\Upsilon }}}}^{{}}$, $\mathbf{P}_{{\mathbf{\tilde{\Upsilon }}}}^{{}}=\mathbf{\tilde{\Upsilon }}{{({{\mathbf{\tilde{\Upsilon }}}^{H}}\mathbf{\tilde{\Upsilon }})}^{-1}}{{\mathbf{\tilde{\Upsilon }}}^{H}}$, $\mathbf{P}_{{\mathbf{\tilde{B}}}}^{\bot }={{\mathbf{I}}_{N}}-\mathbf{P}_{{\mathbf{\tilde{B}}}}^{{}}$, and $\mathbf{P}_{{\mathbf{\tilde{B}}}}^{{}}=\mathbf{\tilde{B}}{{({{\mathbf{\tilde{B}}}^{H}}\mathbf{\tilde{B}})}^{-1}}{{\mathbf{\tilde{B}}}^{H}}$. The two detectors expressed in \eqref{34} and \eqref{35} are termed as the ordinary GLRT to suppress interference (GLRT-I) and 2S-GLRT to suppress interference (2S-GLRT-I), respectively. Additionally, we also consider the Bayesian GLRT to suppress interference (B-GLRT-I) and Bayesian 2S-GLRT to suppress interference (B-2S-GLRT-I) in \cite{Gao2017}, described as
\begin{equation}\label{36}
	{{t}_{\text{B-GLRT-I}}}=\frac{\left| {{\mathbf{I}}_{K}}+{{{\mathbf{\breve{Z}}}}^{H}}\mathbf{P}_{{\mathbf{\breve{\Upsilon }}}}^{\bot }\mathbf{\breve{Z}} \right|}{\left| {{\mathbf{I}}_{K}}+{{{\mathbf{\breve{Z}}}}^{H}}\mathbf{P}_{{\mathbf{\breve{B}}}}^{\bot }\mathbf{\breve{Z}} \right|}\end{equation}
and
\begin{equation}\label{37}
	{{t}_\text{B-2S-GLRT-I}}=\text{tr}\left[ {{{\mathbf{\breve{Z}}}}^{H}}\mathbf{P}_{{\mathbf{\breve{\Upsilon }}}}^{\bot }\mathbf{\breve{\Phi}}{{({{{\mathbf{\breve{\Phi}}}}^{H}}\mathbf{P}_{{\mathbf{\breve{\Upsilon }}}}^{\bot }\mathbf{\breve{\Phi}})}^{-1}}{{{\mathbf{\breve{\Phi}}}}^{H}}\mathbf{P}_{{\mathbf{\breve{\Upsilon }}}}^{\bot }\mathbf{\breve{Z}} \right]\end{equation}
respectively,
For simulated data, we can set signal-to-noise ratio (SNR) and interference-to-noise ratio (INR) as
\begin{equation}\label{38}
	\text{SNR}=\text{tr}({{\mathbf{A}}^{H}}{{\mathbf{\Phi}}^{H}}{{\Sigmabf }^{-1}}\mathbf{{\Phi}A})\end{equation}
and
\begin{equation}\label{39}
	\text{INR}=\text{tr}({{\mathbf{W}}^{H}}{{\mathbf{\Upsilon }}^{H}}{{\Sigmabf }^{-1}}\mathbf{\Upsilon {W}})\end{equation}
the $(i,j)$th element of the scale matrix is $\Sigmabf (i,j)=\sigma^2{{\rho  }^{\text{abs}(i-j)}}$, where $\text{abs}(\cdot )$ denotes the absolute value. Moreover, we let the system dimension as $N=10$ and INR=10 dB. To reduce the complex of computation, we let the probability of false alarm (PFA) as $\text{PFA}={{10}^{-3}}$ for simulated data. ${{10}^{4}}$ Monte Carlo simulations are run to derive the PD for the results of simulation. To get the detection threshold for the predesigned PFA, we run ${100}/{\text{PFA}}$ Monte Carlo simulations.

The real data utilized in experiment are from the IPIX dataset ``$19980223\_170435\_\text{IPIX.mat}$'', which contains 60000 pulse from 34 range bins. We set $\text{PFA}={{10}^{-2}}$ since the quantity of the real data are limited. As we all know, the IPIX dataset are compound-Gaussian. Therefore, first of all, we adopt the pre-processing in \cite{8939391} to the data to make them to be Gaussian. Moreover, for real data, the matrix $\Sigmabf $ in \eqref{38} and \eqref{39} is substituted to the SCM estimated by the whole available data.

\subsection{Simulated Data}

\begin{figure}[!htp]
	\centering
	\subfloat[$\rho=0.9$]{\includegraphics[width=0.4935\linewidth]{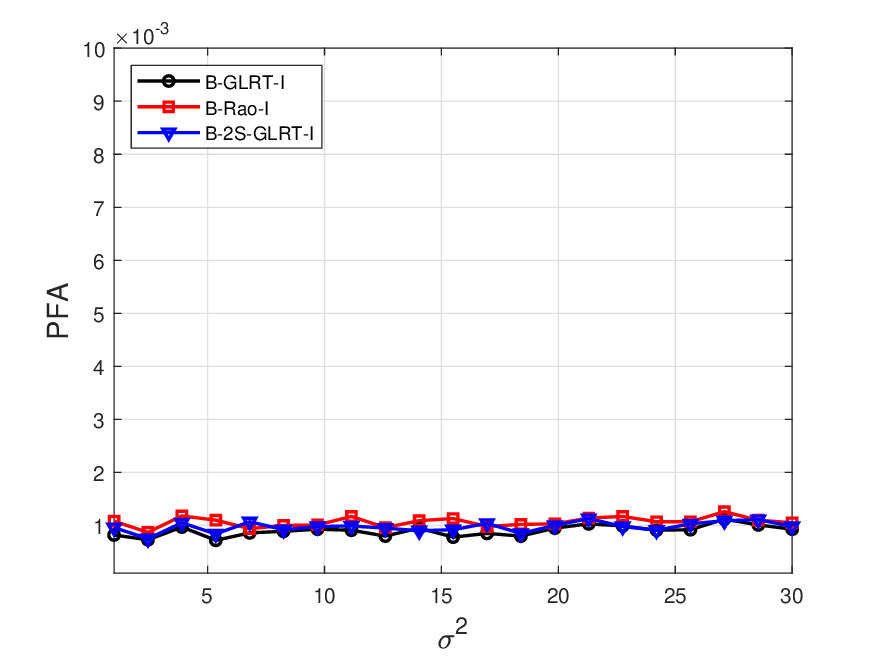}}
	\subfloat[$\sigma^2=1$]{\includegraphics[width=0.4935\linewidth]{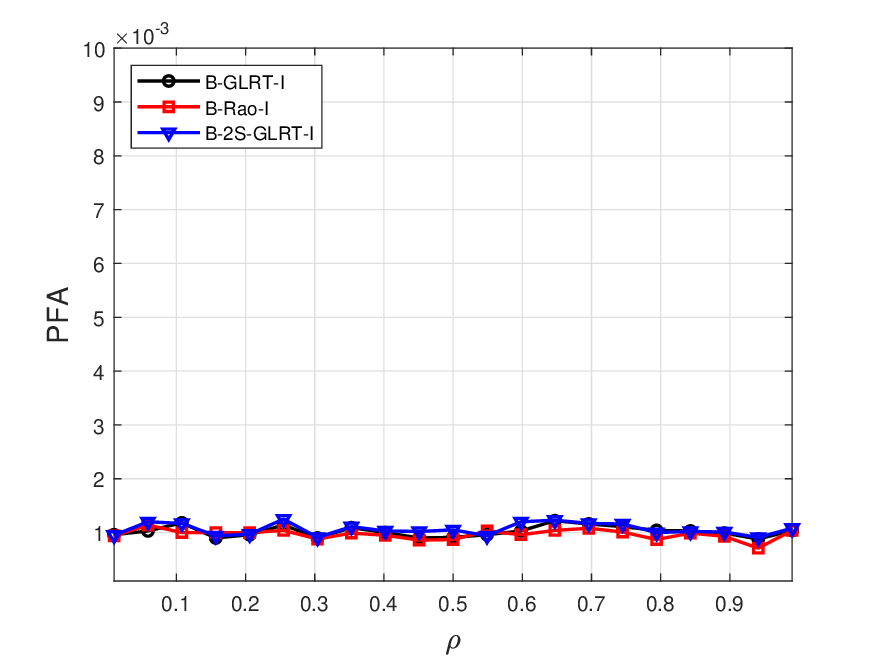}}
	\caption{PFA varies with $\sigma^2$ and $\rho$ for simulated data. $p=7$, $q=3$, $K=4$, and $L=12$.}
	\label{PFA}
\end{figure}
Fig. 1 shows the PFAs of the proposed B-Rao-I and the existing Bayesian detectors B-GLRT-I and B-2S-GLRT-I when $\sigma^2$ and $\rho$ vary. From the results presented in the figure, it can be seen that the PFAs of the Bayesian detector hardly changes with the variation of $\sigma^2$ or $\rho$, indicating the CFAR property. However, it is difficult to provide a rigorous mathematical proof of the CFAR property of the detectors. This is a problem worthy of research in the future.

\begin{figure}[!htp]
	\centering
	\subfloat[$\eta =14$]{\includegraphics[width=0.4935\linewidth]{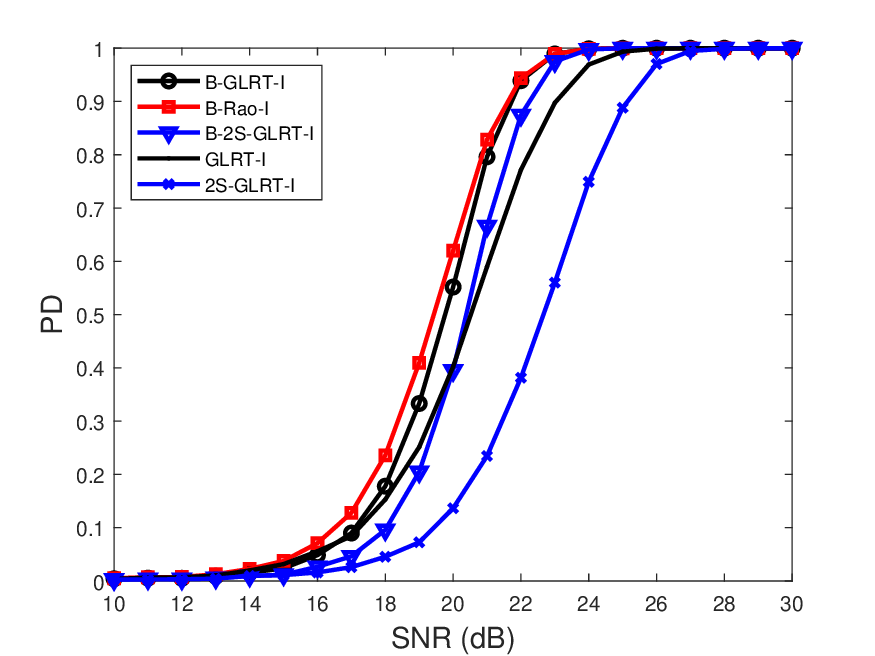}}
	\subfloat[ $\eta =22$]{\includegraphics[width=0.4935\linewidth]{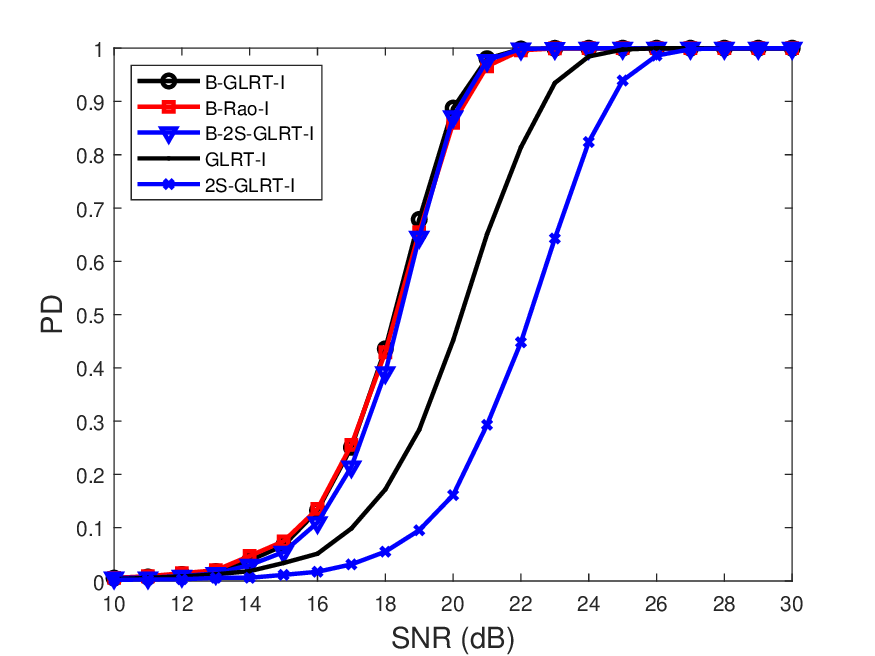}}
	\caption{PD varies with SNR for simulated data. $p=7$, $q=3$, $K=4$, $L=12$, $\sigma^2=1$, and $\rho=0.9$.}
	\label{Fig1}
\end{figure}
Fig. 2 indicates the PDs varies with SNRs with $\eta$ as a variable parameter. We can find that the B-Rao-I detector achieves superior detection performance than other detectors. One plausible explanation for the performance enhancement of the proposed B-Rao-I compared to the B-GLRT-I lies in the Rao test's inherent capability to handle limited training data more effectively. Notably, the Rao criterion was initially proposed in a context with fewer training data \cite{Rao48}.
Moreover, all the Bayesian detectors outperform ordinary detectors, highlighting the efficacy of the Bayesian approach in adaptive detection. Comparing the two sub-figures in Fig. 2 shows that the increase of the DOF $\eta $ leads to the capability enhanced of the Bayesian detectors. This can be attributed to the greater reliability of the \emph{a priori} information in regard to the covariance matrix associated with higher values of $\eta$.

\begin{figure}[!htp]
	\centering
	\subfloat[$p=3$]{\includegraphics[width=0.4935\linewidth]{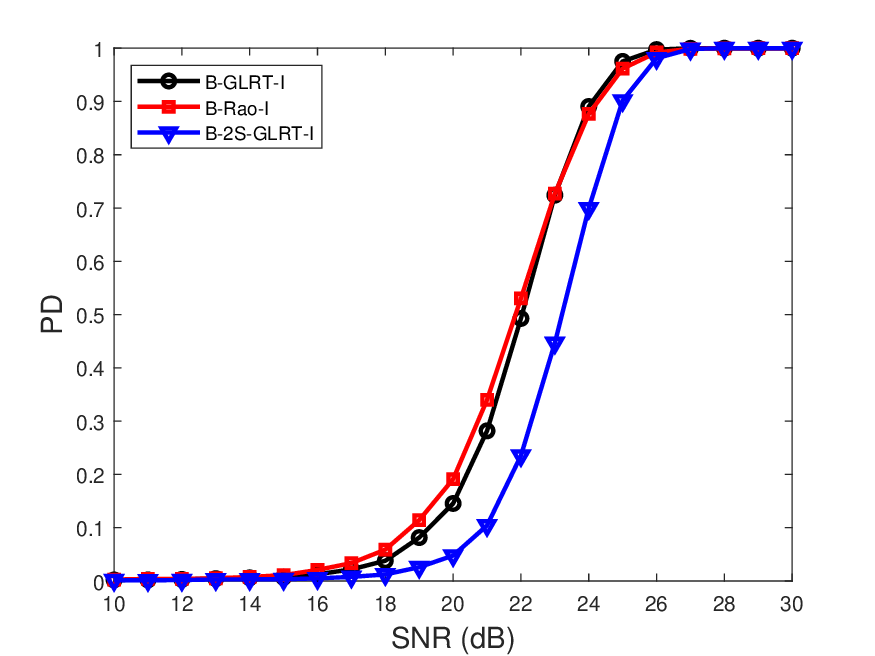}}
	\subfloat[$p=5$]{\includegraphics[width=0.4935\linewidth]{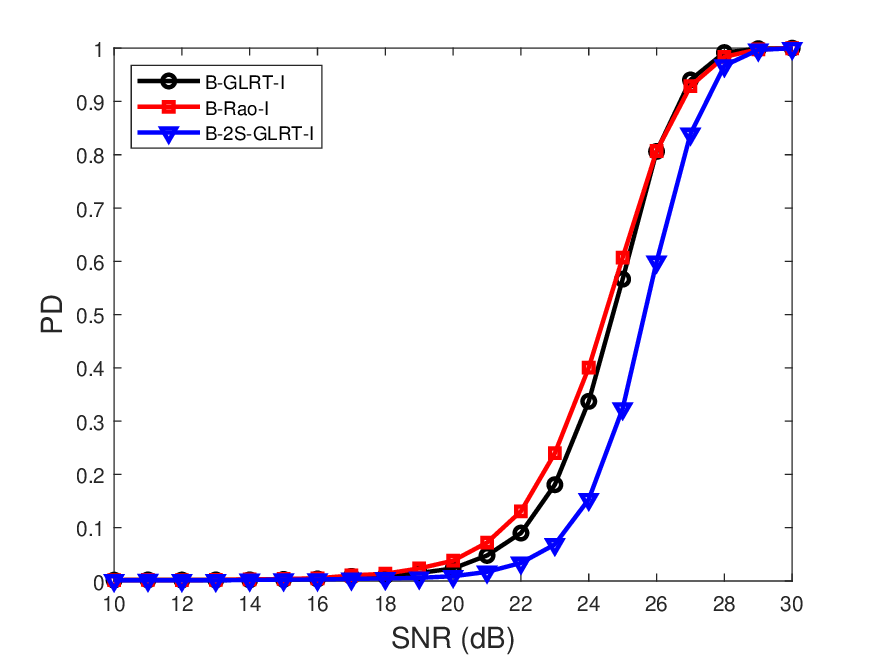}}
	\caption{PD varies with SNR for simulated data. $q=5$, $K=6$, $\eta =14$, $L=8$, $\sigma^2=1$, and $\rho=0.9$.}
	\label{Fig2}
\end{figure}
Fig. 3 presents the PDs across different SNRs with $p$ as a variable parameter. However, the PDs of the ordinary detectors are not shown due to the invalidity of them in such a sample-starved environment. It is evident that for the selected parameters the proposed B-Rao-I can attain roughly the best detection performance. Additionally, The performance of the Bayesian detectors deteriorates when the signal subspace dimension extends. This is due to the dispersion of signal energy, leading to a negatively impact of detection performance.

\begin{figure}[!htp]
	\centering
	\subfloat[$q=1$]{\includegraphics[width=0.4935\linewidth]{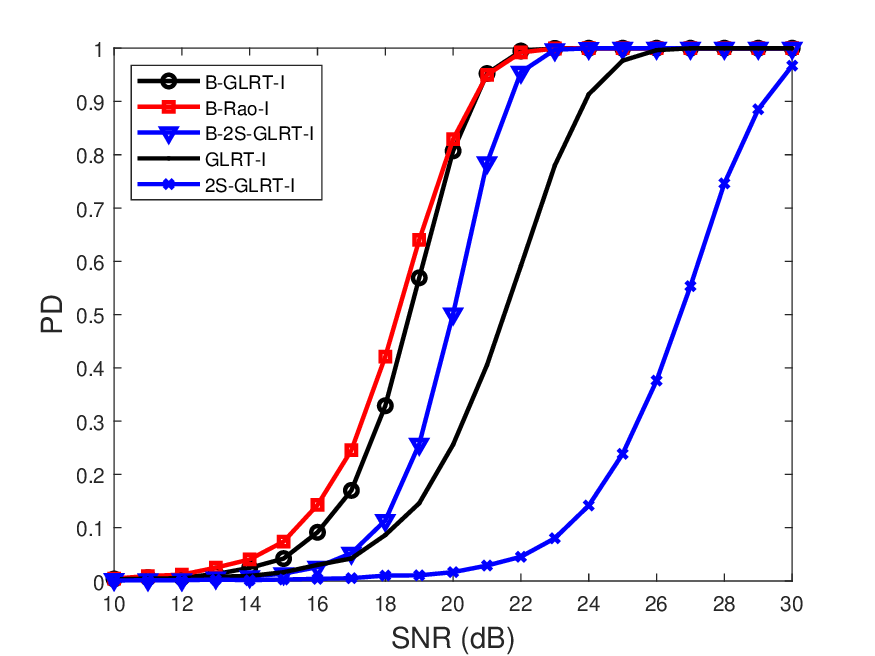}}
	\subfloat[$q=2$]{\includegraphics[width=0.4935\linewidth]{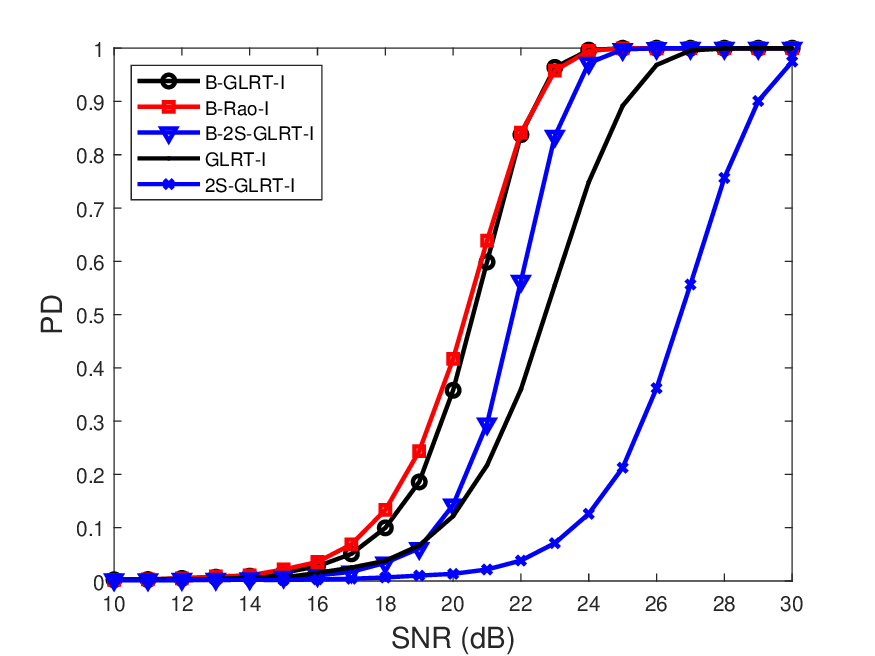}}
	\caption{PD varies with SNR for simulated data. $p=6$, $K=9$, $\eta =14$,  $L=12$, $\sigma^2=1$, and $\rho=0.9$.}
	\label{Fig3}
\end{figure}
Fig. 4 depicts the PDs varies with SNRs with $q$ as a variable parameter. The results demonstrate that the performance of B-Rao-I is superior than the other detectors. Furthermore, it is observed that the PD of a detector diminishes as the interference subspace dimension extends. This decline is a result of the increased leakage of signal energy that is projected into the subspace of interference.

\begin{figure}[!htp]
	\centering
	\subfloat[$L=8$]{\includegraphics[width=0.4935\linewidth]{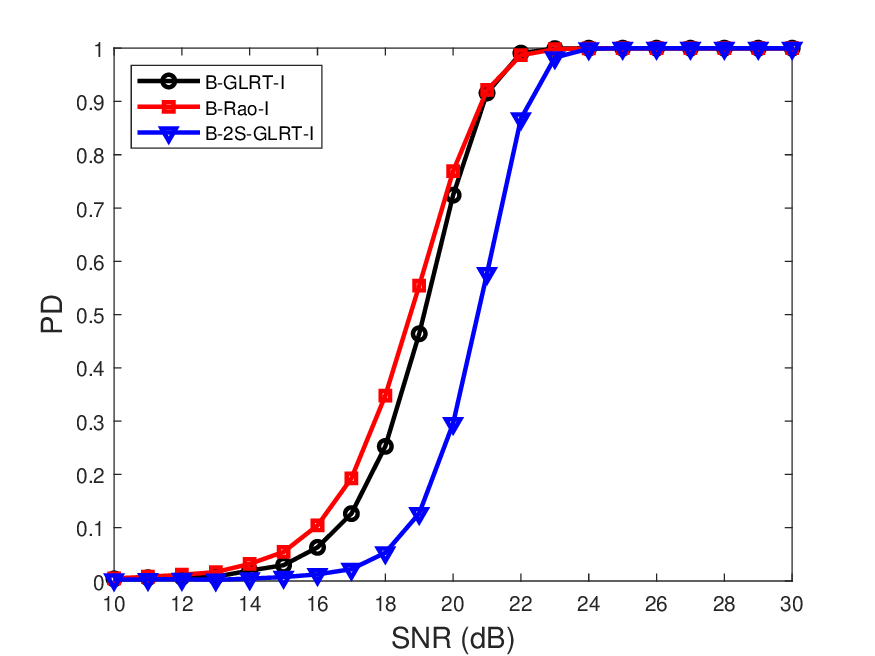}}
	\subfloat[$L=13$]{\includegraphics[width=0.4935\linewidth]{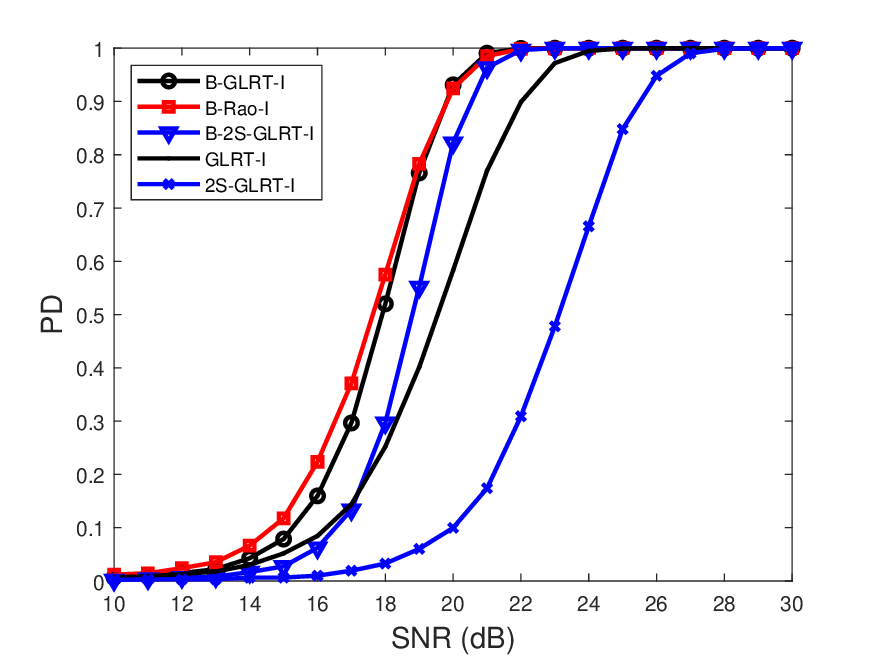}}
	\caption{PD varies with SNR for simulated data. $p=8$, $q=1$, $K=6$, $\eta =14$, $\sigma^2=1$, and $\rho=0.9$.}
	\label{Fig4}
\end{figure}
Fig. 5 shows the PDs across various SNRs with $L$ as a variable parameter. The results demonstrate that the detectors performance enhance as the amount of training data increases. Moreover, the ordinary detectors are invalid because inverse of the covariance matrix is unable to be formed due to too few training data, and hence they are not shown in Fig. 5 (a).

\subsection{Real Data}

\begin{figure}[]
	\centering
	\subfloat[$\eta=14$]{\includegraphics[width=0.4935\linewidth]{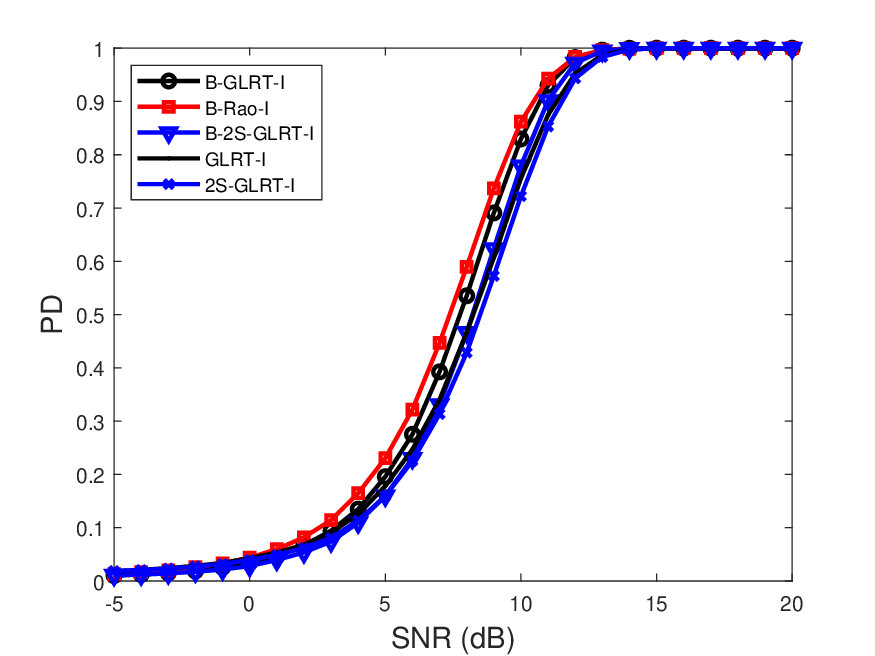}}
	\subfloat[$\eta=22$]{\includegraphics[width=0.4935\linewidth]{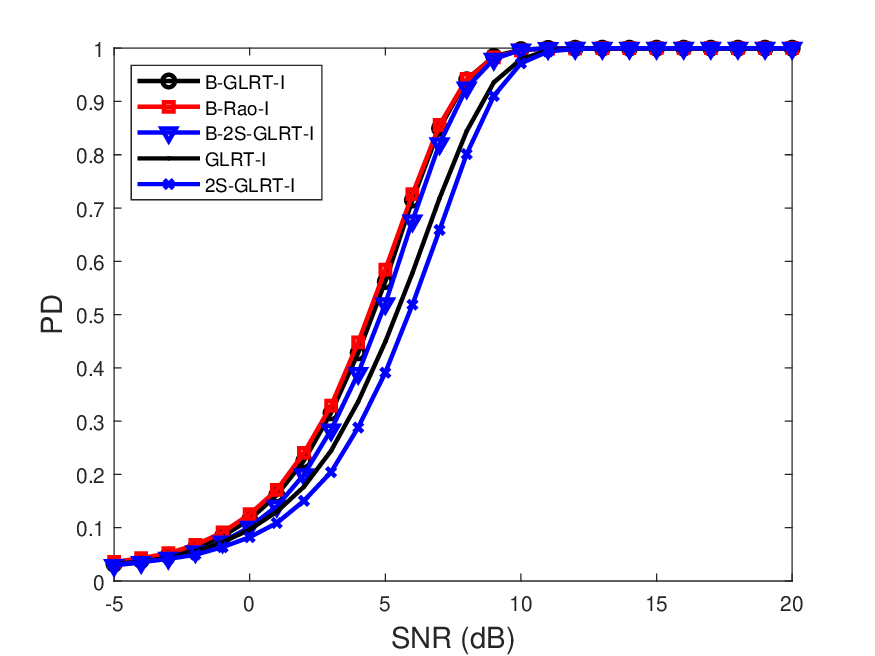}}
	\caption{PD varies with SNR for real data. $p=1$, $q=7$, $K=4$, and $L=12$.}
	\label{Fig5}
\end{figure}

Fig. 6 illustrates the PDs at various SNRs with $\eta $ as a variable parameter. It is indicated that the B-Rao-I offers superior detection performance compared with other detectors. Moreover, The performances of Bayesian detectors improve with the DOF $\eta $ increases.

\begin{figure}[]
	\centering
	\subfloat[$p=1$]{\includegraphics[width=0.4935\linewidth]{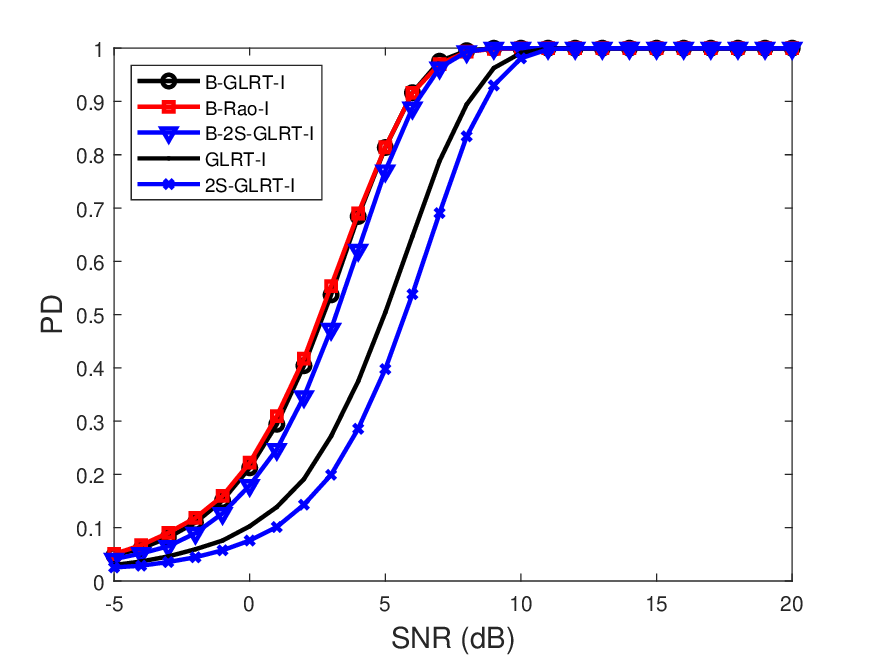}}
	\subfloat[$p=2$]{\includegraphics[width=0.4935\linewidth]{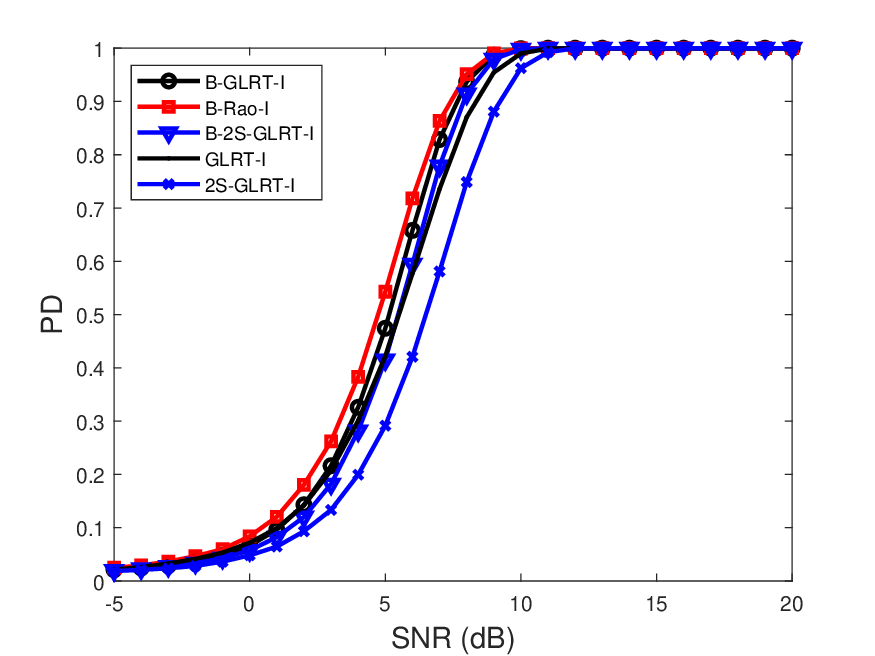}}
	\caption{PD varies with SNR for real data. $q=6$, $K=4$, $\eta =22$, and $L=12$. }
	\label{Fig6}
\end{figure}
Fig. 7 gives the PDs across various SNRs with $p$ as a variable parameter. The graphs reveal that the Bayesian detectors can possess higher PDs than the ordinary detectors. Additionally, the PDs for the whole detectors exhibit a decline as the dimension of the signal subspace $p$ extends.

\begin{figure}[]
	\centering
	\subfloat[$q=4$]{\includegraphics[width=0.4935\linewidth]{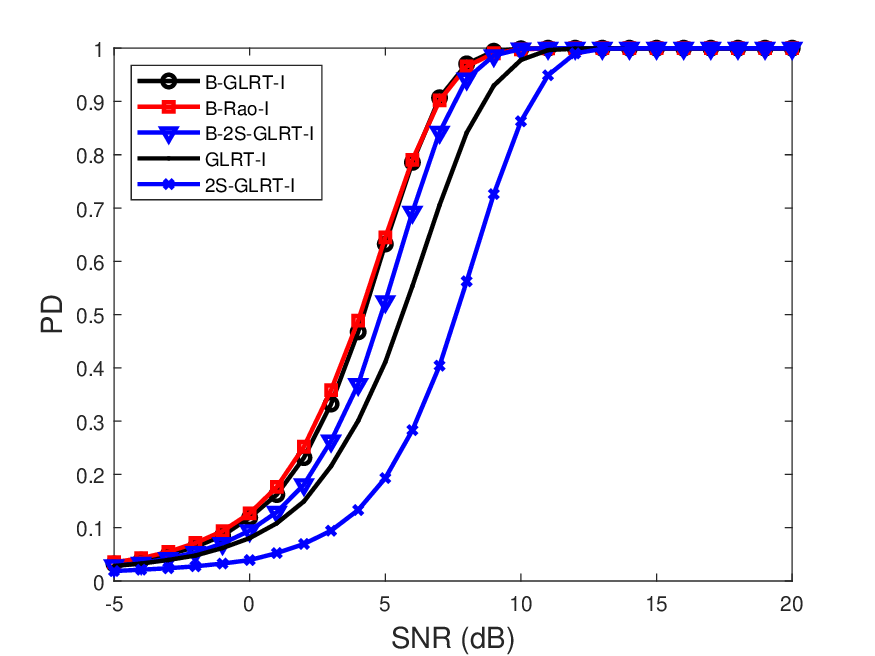}}
	\subfloat[$q=7$]{\includegraphics[width=0.4935\linewidth]{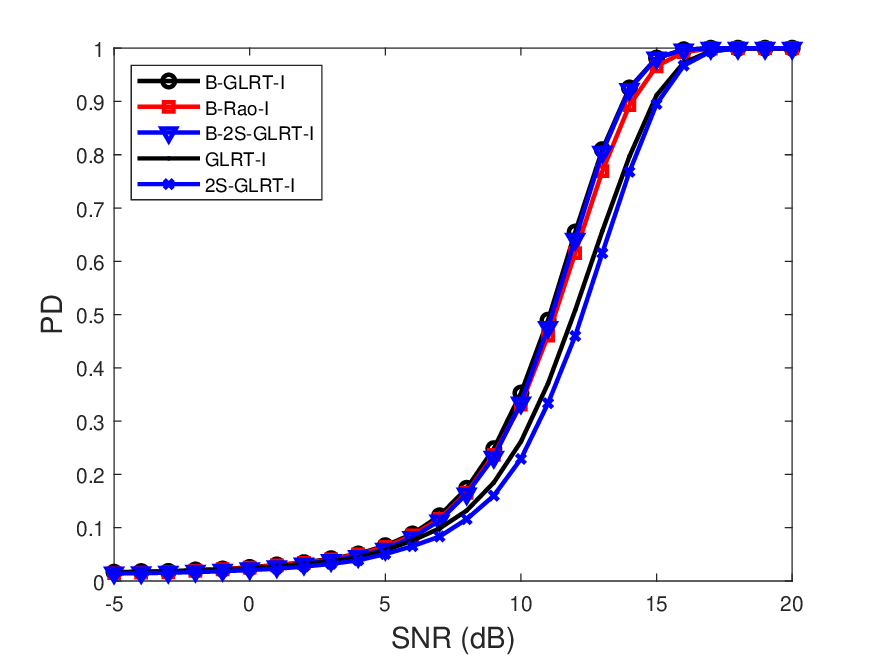}}
	\caption{PD varies with SNR for real data.  $p=2$, $K=4$, $\eta =14$, and $L=12$. }
	\label{Fig7}
\end{figure}
Fig. 8 illustrates the PDs across different SNRs with $q$ as a variable parameter. The results reveal that the detectors' detectability all decrease when the interference subspace dimension $q$ extends. As pointed out above, this decline is owing to the leakage of signal energy that projected into the subspace of interference.

\begin{figure}[]
	\centering
	\subfloat[$L=12$]{\includegraphics[width=0.4935\linewidth]{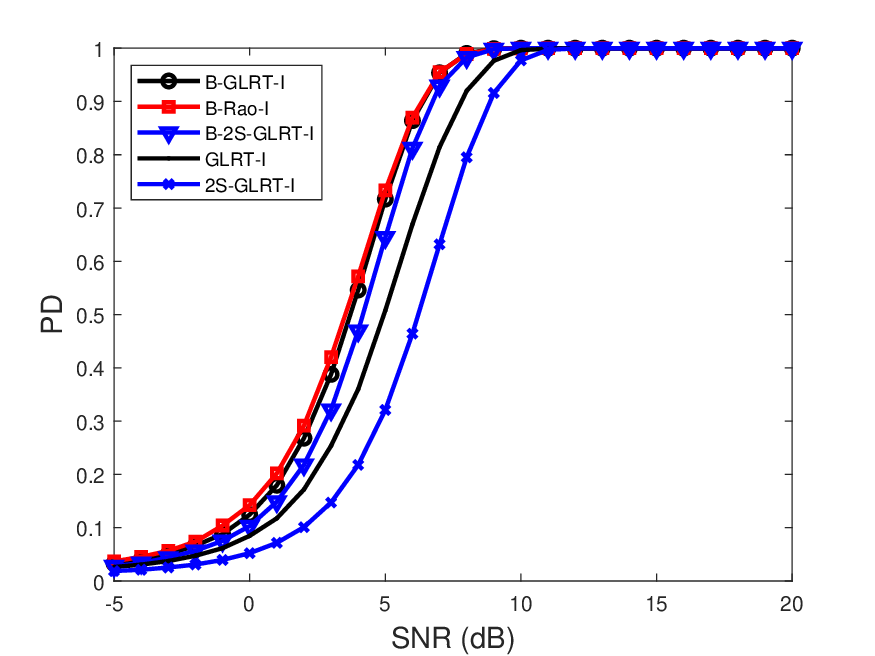}}
	\subfloat[$L=17$]{\includegraphics[width=0.4935\linewidth]{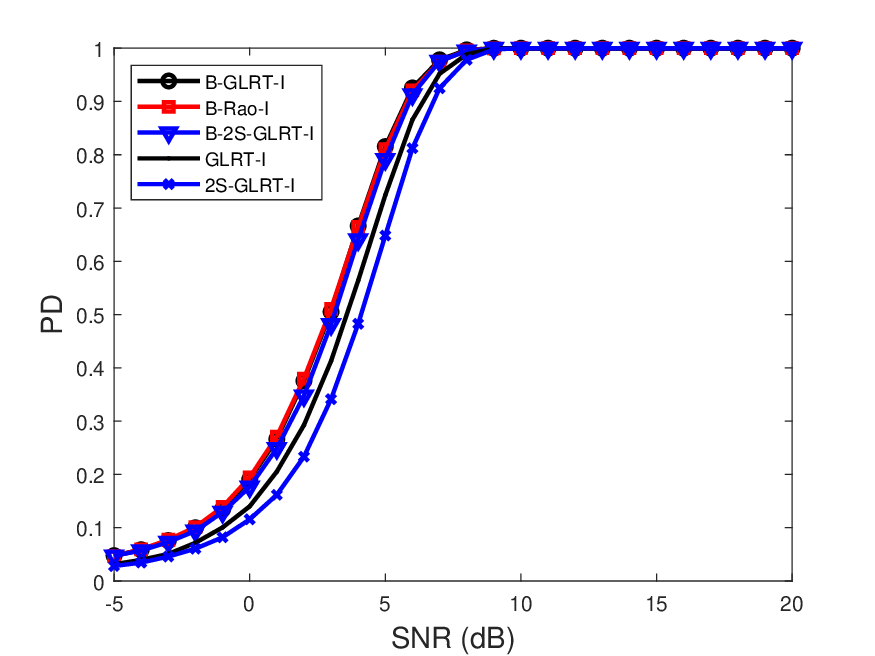}}
	\caption{PD varies with SNR for real data.  $p=3$, $q=5$, $K=4$, and $\eta =14$. }
	\label{Fig8}
\end{figure}
Fig. 9 gives the PDs varies with SNRs with $L$ as a variable parameter. The graph demonstrates that the performance of the whole detectors have enhanced as the quantity of training sanples $L$ increase. The enhancement is a result of the more precise estimation of the SCM. In addition, the degree of improvement of the conventional detectors is a little higher than that of the Bayesian detectors.

Before closing this section, we would like to briefly discuss the computational complexities of the B-Rao-I and B-GLRT-I.
The B-Rao-I, derives its primary computational complexity from the inversion of an $N\times N$ dimensional matrix, thus resulting in a computational complexity of $O(N^3)$. The B-GLRT-I has its main computational complexity from the inversion of an $N\times N$ dimensional matrix as well as the determinant computation of a $K\times K$ dimensional matrix.
For matrices of the same dimension, since matrix inversion and determinant computation share the same level of complexity, the computational complexity of the B-GLRT-I detector is $\max\{O(N^3), O(K^3)\}$. Therefore, it can be concluded that the proposed B-Rao-I detector has a computational complexity that is comparable to or less than that of the B-GLRT-I.
\section{Conclusion}
This paper addressed the challenge of detecting a range-spread target in subspace interference and noise when there was limited training data. To solve this problem, we developed the Bayesian detector B-Rao-I. Through both simulated and real data experiments, we demonstrated that B-Rao-I outperformed existing detectors. Our investigation also revealed that the effectiveness of Bayesian detectors was enhanced by reducing the dimensionality of the signal and interference subspaces, increasing the DOFs of the inverse Wishart distribution, and increasing the amount of training data. Potential directions involve generalizing the proposed B-Rao-I to other types of clutter models and analyzing the sensitivity of the detector to the assumption on data statistics.

\section*{APPENDIX. {The B-2S-GLRT-I is equivalent to the Bayesian Wald Test}}

In this appendix, we show that the B-2S-GLRT-I can also be derived based on Wald Test, which is given by \cite{Liu2014}
\begin{equation}\label{40}
	{{t}_\text{Wald}}={{({{\mathbf{\hat{\Thetabf }}}_{{{\text{r}}_{\text{1}}}}}-{{\mathbf{\Thetabf }}_{{{\text{r}}_{\text{0}}}}})}^{H}}{{\left\{ {{[{{\mathbf{F}}^{-1}}({{{\mathbf{\hat{\Thetabf }}}}_{1}})]}_{{{\mathbf{\Thetabf }}_{\text{r}}},{{\mathbf{\Thetabf }}_{\text{r}}}}} \right\}}^{-1}}({{\mathbf{\hat{\Thetabf }}}_{{{\text{r}}_{\text{1}}}}}-{{\mathbf{\Thetabf }}_{{{\text{r}}_{\text{0}}}}})\end{equation}
where ${{\mathbf{\hat{\Thetabf }}}_{{{\text{r}}_{1}}}}$ represents the MLE of ${{\mathbf{\Thetabf }}_{\text{r}}}$ under ${\text{H}_{1}}$, ${{\mathbf{\Thetabf }}_{{{\text{r}}_{0}}}}$ represents ${{\mathbf{\Thetabf }}_{\text{r}}}$ under ${\text{H}_{0}}$.
Nulling \eqref{15} w.r.t. $\mathbf{C}$, we can easily get the MLE of $\mathbf{C}$ for given $\mathbf{R}$ as
\begin{equation}\label{41}
	\mathbf{\hat{C}}={{({{\mathbf{B}}^{H}}{{\mathbf{R}}^{-1}}\mathbf{B})}^{-1}}{{\mathbf{B}}^{H}}{{\mathbf{R}}^{-1}}\mathbf{Z}\end{equation}
The MLE of $\mathbf{A}$ is the first $p$ rows of \eqref{41}. As the method described by \eqref{19} in \cite{9591291}, we can get the MLE of $\mathbf{A}$ from \eqref{41} as
\begin{equation}\label{42}
	\mathbf{\hat{A}}={{({{\mathbf{\bar{\Phi}}}^{H}}\mathbf{P}_{{\mathbf{\bar{\Upsilon }}}}^{\bot }\mathbf{\bar{\Phi}})}^{-1}}{{\mathbf{\bar{\Phi}}}^{H}}\mathbf{P}_{{\mathbf{\bar{\Upsilon }}}}^{\bot }\mathbf{\bar{Z}}\end{equation}
From \eqref{13} we have
\begin{equation}\label{43}
	{{\left\{ {{[{{\mathbf{F}}^{-1}}(\mathbf{\Thetabf })]}_{{{\mathbf{\Thetabf }}_{\text{r}}},{{\mathbf{\Thetabf }}_{\text{r}}}}} \right\}}^{-1}}={{\mathbf{I}}_{K}}\otimes ({{\mathbf{\bar{\Phi}}}^{H}}\mathbf{P}_{{\mathbf{\bar{\Upsilon }}}}^{\bot }\mathbf{\bar{\Phi}})\end{equation}
Then substituting \eqref{42} and \eqref{43} into \eqref{40} results in the Wald test for given $\mathbf{R}$ as
\begin{equation}\label{44}
	\begin{aligned}
		{{t}_{{\text{Wald}_{\mathbf{R}}}}}=&\text{ve}{{\text{c}}^{H}}\left[ {{({{{\mathbf{\bar{\Phi}}}}^{H}}\mathbf{P}_{{\mathbf{\bar{\Upsilon }}}}^{\bot }\mathbf{\bar{\Phi}})}^{-1}}{{{\mathbf{\bar{\Phi}}}}^{H}}\mathbf{P}_{{\mathbf{\bar{\Upsilon }}}}^{\bot }\mathbf{\bar{Z}} \right]({{\mathbf{I}}_{K}}\otimes {{{\mathbf{\bar{\Phi}}}}^{H}}\mathbf{P}_{{\mathbf{\bar{\Upsilon }}}}^{\bot }\mathbf{\bar{\Phi}})
		\text{vec}\left[ {{({{{\mathbf{\bar{\Phi}}}}^{H}}\mathbf{P}_{{\mathbf{\bar{\Upsilon }}}}^{\bot }\mathbf{\bar{\Phi}})}^{-1}}{{{\mathbf{\bar{\Phi}}}}^{H}}\mathbf{P}_{{\mathbf{\bar{\Upsilon }}}}^{\bot }\mathbf{\bar{Z}} \right] \\
		=& \text{ve}{{\text{c}}^{H}}\left[ {{({{{\mathbf{\bar{\Phi}}}}^{H}}\mathbf{P}_{{\mathbf{\bar{\Upsilon }}}}^{\bot }\mathbf{\bar{\Phi}})}^{-1}}{{{\mathbf{\bar{\Phi}}}}^{H}}\mathbf{P}_{{\mathbf{\bar{\Upsilon }}}}^{\bot }\mathbf{\bar{Z}} \right]
		\text{vec}\left[ {{{\mathbf{\bar{\Phi}}}}^{H}}\mathbf{P}_{{\mathbf{\bar{\Upsilon }}}}^{\bot }\mathbf{\bar{\Phi}}{{({{{\mathbf{\bar{\Phi}}}}^{H}}\mathbf{P}_{{\mathbf{\bar{\Upsilon }}}}^{\bot }\mathbf{\bar{\Phi}})}^{-1}}{{{\mathbf{\bar{\Phi}}}}^{H}}\mathbf{P}_{{\mathbf{\bar{\Upsilon }}}}^{\bot }\mathbf{\bar{Z}} \right] \\
		=& \text{tr}\left[ {{{\mathbf{\bar{Z}}}}^{H}}\mathbf{P}_{{\mathbf{\bar{\Upsilon }}}}^{\bot }\mathbf{\bar{\Phi}}{{({{{\mathbf{\bar{\Phi}}}}^{H}}\mathbf{P}_{{\mathbf{\bar{\Upsilon }}}}^{\bot }\mathbf{\bar{\Phi}})}^{-1}}{{{\mathbf{\bar{\Phi}}}}^{H}}\mathbf{P}_{{\mathbf{\bar{\Upsilon }}}}^{\bot }\mathbf{\bar{Z}} \right]
\end{aligned}\end{equation}
which shares the same form as \eqref{17}. Hence, we can also rewrite \eqref{44} as \eqref{18}, repeated below
\begin{equation}\label{45}
	{{t}_{\text{Wal}{{\text{d}}_{\mathbf{R}}}}}=\text{tr}({{\Deltabf }^{H}}\Lambdabf \Deltabf )
\end{equation}
Next, We aim to derive the MAP of $\mathbf{R}$ under ${{\text{H}}_{1}}$. One can verify that
\begin{equation}\label{46}
	\begin{aligned}
		\left| {{\mathbf{Z}}_{1}}\mathbf{Z}_{1}^{H}+\mathbf{S}+\eta \Sigmabf  \right|=&\left| (\mathbf{Z}-\mathbf{BC}){{(\mathbf{Z}-\mathbf{BC})}^{H}}+\mathbf{S}+\eta \Sigmabf  \right| \\
		=&\left| \mathbf{S}+\eta \Sigmabf  \right|\cdot \left| {{\mathbf{I}}_{K}}+{{(\mathbf{Z}-\mathbf{BC})}^{H}}{{(\mathbf{S}+\eta \Sigmabf )}^{-1}}(\mathbf{Z}-\mathbf{BC}) \right| \\
		=&\left| \mathbf{S}+\eta \Sigmabf  \right|\cdot \left| {{\mathbf{I}}_{K}}+{{(\mathbf{\breve{Z}}-\mathbf{\breve{B}C})}^{H}}(\mathbf{\breve{Z}}-\mathbf{\breve{B}C}) \right|
\end{aligned}\end{equation}
where $\mathbf{\breve{B}}={{(\mathbf{S}+\eta \Sigmabf )}^{-1/2}}\mathbf{B}$. Setting the derivative of \eqref{46} w.r.t. $\mathbf{C}$ to be zero results in the MAP of $\mathbf{C}$ as
\begin{equation}\label{47}
	\mathbf{\hat{C}}={{({{\mathbf{\breve{B}}}^{H}}\mathbf{\breve{B}})}^{-1}}{{\mathbf{\breve{B}}}^{H}}\mathbf{\breve{Z}}\end{equation}
Setting the deriveative of \eqref{15} w.r.t. $\mathbf{R}$ to be zero under ${{\text{H}}_{1}}$, we get the MAP of $\mathbf{R}$ under hypothesis ${{\text{H}}_{1}}$ for given $\mathbf{A}$ and $\mathbf{W}$ as
\begin{equation}\label{48}
	{{\mathbf{\hat{R}}}_{1}}=\frac{(\mathbf{Z}-\mathbf{BC}){{(\mathbf{Z}-\mathbf{BC})}^{H}}+\mathbf{S}+\eta \Sigmabf }{\eta +N+L+K}\end{equation}
Substituting \eqref{47} in \eqref{48} leads to
\begin{equation}\label{49}
	{{\mathbf{\hat{R}}}_{1}}={{(\mathbf{S}+\eta \Sigmabf )}^{1/2}}\frac{\mathbf{P}_{{\mathbf{\breve{B}}}}^{\bot }\mathbf{\breve{Z}}{{{\mathbf{\breve{Z}}}}^{H}} \mathbf{P}_{{\mathbf{\breve{B}}}}^{\bot }+{{\mathbf{I}}_{N}}}{\eta +N+L+K}{{(\mathbf{S}+\eta \Sigmabf )}^{1/2}}\end{equation}
where $\mathbf{P}_{{\mathbf{\breve{B}}}}^{\bot }={{\mathbf{I}}_{N}}-{{\mathbf{P}}_{{\mathbf{\breve{B}}}}}$ and ${{\mathbf{P}}_{{\mathbf{\breve{B}}}}}=\mathbf{\breve{B}}{{({{\mathbf {\breve{B}}}^{H}}\mathbf{\breve{B}})}^{-1}}{{\mathbf {\breve{B}}}^{H}}$. Performing the matrix inversion to \eqref{49} leads to
\begin{equation}\label{50}
	\begin{aligned}
		\mathbf{\hat{R}}_{1}^{-1}=&{{\alpha }}{{(\mathbf{S}+\eta \Sigmabf )}^{-1/2}}\left[ {{\mathbf{I}}_{N}}-\mathbf{P}_{{\mathbf{\breve{B}}}}^{\bot }\mathbf{\breve{Z}}{{\left( {{\mathbf{I}}_{K}}+{{{\mathbf{\breve{Z}}}}^{H}}\mathbf{P}_{{\mathbf{\breve{B}}}}^{\bot }\mathbf{\breve{Z}} \right)}^{-1}}{{{\mathbf{\breve{Z}}}}^{H}}\mathbf{P}_{{\mathbf{\breve{B}}}}^{\bot } \right]
		{{(\mathbf{S}+\eta \Sigmabf )}^{-1/2}} \\
	\end{aligned}
\end{equation}
Using \eqref{50}, we can easily obtain the following two equations
\begin{equation}\label{51}
	\mathbf{\hat{R}}_{1}^{-1}\mathbf{\Phi}={{\alpha }}{{(\mathbf{S}+\eta \Sigmabf )}^{-1}}\mathbf{\Phi}\end{equation}
and
\begin{equation}\label{52}
	\mathbf{\hat{R}}_{1}^{-1}\mathbf{\Upsilon }={{\alpha }}{{(\mathbf{S}+\eta \Sigmabf )}^{-1}}\mathbf{\Upsilon }\end{equation}
Substituting \eqref{51} and \eqref{52} into \eqref{19} and \eqref{20}, and then inserting the results back into \eqref{45} while disregarding the constant term, we can get the ultimate Wald test, which is isovalent to the B-2S-GLRT-I in \eqref{37}.

\Acknowledgements{This work was supported by National Natural Science Foundation of China (Grant Nos. 62071482, 62471485, and 62471450), the Hubei Provincial Natural Science Foundation of China (Grant Nos. 2025AFB873, 2023AFA035, 2024BAA005), Natural Science
Foundation of Anhui Province (Grant No. 2208085J17), and the Central Government Guided Local Funds for Science and Technology Development (2024CSA080).}



%
\bibliographystyle{ieeetr}
\bibliography{BaysRao.bib}
\end{document}